\documentclass[prb,twocolumn,showpacs,superscriptaddress,floatfix]{revtex4}
\usepackage{graphicx,color,amsfonts,amsmath}

\newcommand{\rmd}{\mathrm{d}}
\newcommand{\rmc}{\mathrm{c}}
\newcommand{\rms}{\mathrm{s}}
\newcommand{\rmcs}{{\rmc(\rms)}}
\newcommand{\rmi}{\mathrm{i}}

\newcommand{\rmr}{\mathrm{R}}
\newcommand{\rml}{\mathrm{L}}
\newcommand{\rmph}{\mathrm{ph}}
\newcommand{\rmpp}{\mathrm{pp}}

\newcommand{\cc}{c^\dagger}
\newcommand{\ac}{c^{\phantom\dagger}}
\newcommand{\cpsi}{\psi^\dagger}
\newcommand{\apsi}{\psi^{\phantom\dagger}}

\newcommand{\bS}{\boldsymbol S}

\newcommand{\pal}{\partial_l}

\begin{document}

\title{Quantum wire networks with local $\mathbb Z_2$ symmetry}

\author{K. Kazymyrenko}
\email{kkyrylo@lpthe.jussieu.fr}
\affiliation{Laboratoire de Physique Th\'eorique et Hautes Energies,\\
CNRS UMR 7589, Universit\'e Paris VI et VII,\\
4, Place Jussieu, 75252 Paris Cedex 05, France}
\author{S. Dusuel}
\email{sdusuel@thp.uni-koeln.de}
\affiliation{Institut f\"ur Theoretische Physik, Universit\"at zu K\"oln,
Z\"ulpicher Str. 77, 50937 K\"oln, Germany}
\author{B. Dou\c{c}ot}
\email{doucot@lpthe.jussieu.fr}
\affiliation{Laboratoire de Physique Th\'eorique et Hautes Energies,\\
CNRS UMR 7589, Universit\'e Paris VI et VII,\\
4, Place Jussieu, 75252 Paris Cedex 05, France}

\begin{abstract}
For a large class of networks made of connected loops, in the presence of
an external magnetic field of half flux quantum per loop, we show the
existence of a large local symmetry group, generated by simultaneous flips
of the electronic current in all the loops adjacent to a given node.
Using an ultra-localized single particle basis adapted to this local
$\mathbb Z_2$ symmetry, we show that it is preserved by a large class of
interaction potentials. As a main physical consequence, the only allowed
tunneling processes in such networks are induced by electron-electron interactions
and involve a simultaneous hop of two electrons.
Using a mean-field picture and then a more systematic renormalization-group
treatment, we show that these pair hopping processes do not generate
a superconducting instability, but they destroy the Luttinger liquid behavior
in the links, giving rise at low energy to a strongly correlated spin-density-wave state.

\end{abstract}

\maketitle


\section{Introduction}

For the past two decades, transport properties of quantum wires have received
a lot of attention~\cite{Timp91,Washburn92}. Besides metallic systems, two dimensional
electron gases induced in GaAs/AlGaAs heterostructures have displayed a rich
variety of quantum effects such as Aharonov-Bohm resistance oscillations in
a ring geometry~\cite{Timp87,Pedersen00}, and persistent currents~\cite{Mailly93,Rabaud01}.
More recently, coherent Aharonov-Bohm
oscillations have been measured in ballistic arrays with the dice lattice 
geometry~\cite{Naud01}, in agreement with the predictions of simple models
for non-interacting electrons~\cite{Vidal98}.
The experimental visibility of conductance oscillations with the full flux quantum
\mbox{$\Phi_{0}=h/e$} periodicity in large structures with a few thousand loops
has been attributed to the presence of Aharonov-Bohm cages which localize the motion
of electrons when the external flux per elementary loop is half-integer in units
of $\Phi_{0}$. 

After these first investigations, a natural issue is to understand interaction 
effects on this geometrical localization phenomenon. For two electrons, it was shown
that interaction induces some correlated hopping processes, at the origin of
some delocalized states in which the two electrons always remain close 
from each other~\cite{Vidal00,Vidal01}. Remarkably, this trend appears for any
sign of interaction. Of course, for a repulsive interaction, these extended 
states lie above the ground-state sector, obtained when the two electrons
are individually localized in distinct remote cages. But this raises the question 
of what happens at finite density, where it is no longer possible to put all 
conduction electrons in separate cages. Two electrons initially in the same cage
will be able to move together, so this may generate some kind of metallic state.

An interacting system with a finite particle density was first studied in its
{\em bosonic} version in a one-dimensional geometry showing Aharonov-Bohm cages,
namely a chain of rhombi~\cite{Doucot02}. This model has been adapted into a realistic
description of a Josephson junction array with the same geometry recently~\cite{Protopopov04}.
There, it was found that no single boson condensate can exist, but instead a
2-boson condensate is stabilized at low energy. This peculiar phenomenon is connected
to the single particle localization in Aharonov-Bohm cages, but in the presence of interaction,
it requires a stronger property, identified as a local symmetry based on the $\mathbb Z_2$ 
group. This symmetry implies that the parity of the total particle number within any
Aharonov-Bohm cage is conserved. If this symmetry remains unbroken, it prevents the formation 
of any single boson condensate. A two-dimensional version exhibiting this
local $\mathbb Z_2$ symmetry was also analyzed~\cite{Ioffe02}, in the context
of topologically protected quantum memories. The crucial role of a local symmetry
is well illustrated by the example of the dice lattice, which sustains Aharonov-Bohm
cages, but no local symmetry in the presence of interactions.  
There, it was found that the external magnetic field corresponding to $\Phi_{0}/2$ per loop
imposes a vortex lattice which allows for some zero-energy line defects, but the 
single particle condensate exhibits a finite phase stiffness~\cite{Korshunov01,Korshunov04,Korshunov05}.

The fermionic case at finite density is the subject of the present article.
Motivated by ongoing experiments~\cite{Dufouleur} on networks designed on a
2D electron system at a GaAs/GaAlAs interface, we shall study quantum wire
networks, which geometry is compatible with a local $\mathbb Z_2$ symmetry 
in the presence of a magnetic field of half flux quantum per elementary loop.
Such networks will be called $\mathbb Z_2$ networks, and two examples are 
shown on Fig.~\ref{Z2network} below. 
\begin{figure}[h]
\includegraphics[width=2.7in]{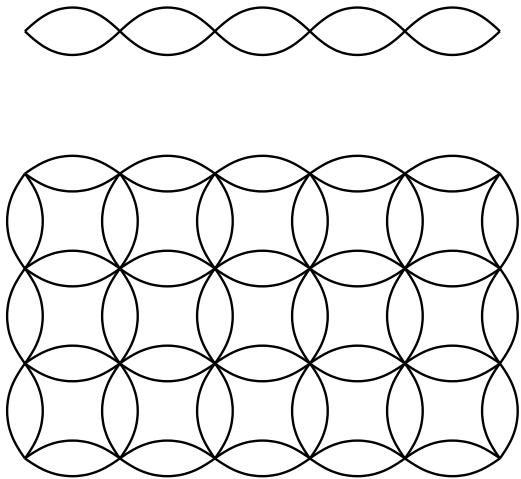}
\caption{This picture presents two examples of $\mathbb Z_2$ networks: a chain of loops
(upper part), and a square lattice of loops (lower part).
All these networks are obtained from any planar lattice after replacing each
link by a loop. To get a local $\mathbb Z_2$ symmetry, each loop is required to
be perfectly invariant under the reflection through the long axis defined by its
two adjacent nodes.}
\label{Z2network}
\end{figure}
In section~\ref{noninteracting}, we define precisely this class of networks,
construct the unitary operators associated to the local $\mathbb Z_2$ symmetry group,
and consider their irreducible representations in the Hilbert space of a single
particle. We show that all the common eigenstates of all the local generators
of this large Abelian group are strongly localized within Aharonov-Bohm cages
involving only a finite number of elementary loops.
Section~\ref{interactioneffects} first discusses the matrix elements of a two-body
interaction in this localized basis adapted to the $\mathbb Z_2$ symmetry.
This shows that a large class of pair interaction potentials (including the point-like
Hubbard interaction) is compatible with this symmetry. We then show that, at least
in a mean-field approach, the particle pair hopping processes generated by 
a repulsive interaction do not lead to superconductivity. Starting from simple order of magnitude 
estimates adapted to semi-conductor quantum wire networks, we argue 
that a more refined renormalization group approach (RG) is required, given the strength
of the Coulomb interaction for typical electronic densities.
Finally, section~\ref{RG} proposes a simplified $\mathbb Z_2$ symmetric model of coupled 
Aharonov-Bohm cages which is analyzed with the RG method using the one-loop approximation.
This confirms the irrelevance of the electron pair hopping processes in the low-energy limit,
and establishes a tendency for the system to reach a strong coupling fixed point of a
spin density wave character. So the combination of local $\mathbb Z_2$ symmetry
and electron-electron repulsion destroys the Luttinger liquid in the quantum wires
which constitute the network, leaving room only for a strongly correlated liquid
with mostly gapped excitations. Finally, a brief appendix shows how this picture
fits with a simple mean-field description.


\section{Local $\mathbb Z_2$ symmetry for a class of quantum networks}
\label{noninteracting}

 \subsection{Basic construction}

In this paper, we shall consider networks of quantum wires (called $\mathbb Z_2$ networks),
for which the elementary links connecting two nearest-neighbor nodes are doubled.
In other words, such networks may be obtained from any planar graph, after replacing links
by loops. Examples of this construction are shown on Fig.~\ref{Z2network}.
The key assumption we shall make
is that all these loops exhibit a perfect reflection symmetry around the axis joining the two nodes
connected by the loop. This requirement imposes some rather severe constraints on the
fabrication of such structures, in terms of the network geometry, and of the control of various
types of disorder. However, as discussed in the Introduction, experimental observation of the
Aharonov-Bohm cage effect in GaAs quantum wire networks~\cite{Naud01} suggests that present
technology is able to meet these constraints. On a more theoretical side, numerical investigations
on disordered tight-binding versions of such networks have shown that the effects we shall
discuss survive after a small but sizeable disorder is introduced\cite{Vidal01}. At a purely geometric level,
these structures exhibit, besides their usual crystalline symmetry groups, a large symmetry
generated by local flips of elementary loops around their central axis. We are mostly
addressing here the quantum-mechanical effects of such a symmetry in the presence of a 
constant and uniform external magnetic field, corresponding to half-flux quantum per loop.

As usual, quantum mechanics is affected by the fact that the vector potential used to describe the
magnetic field is not invariant under these local operations. In the case of the translational
symmetry, this leads to the non-commutation of magnetic translations and to Landau level
quantization. In the present case, the striking consequence is the appearance of a dramatic localization
phenomenon, where all single particle energy eigenstates are confined inside a finite 
set of loops~\cite{Vidal98}. 

To illustrate the idea, it is useful to consider first a single loop, with strictly
one-dimensional motion along a coordinate $x$. If $l$ denotes the perimeter of this loop,
we have to impose the periodicity condition: \mbox{$\psi(l)=\psi(0)$}. The stationary
Schr\"odinger equation reads: 
\[
E\psi(x)=\frac{1}{2m}(\hat p+ eA)^2\psi(x)=\frac{1}{2m}(-i\hbar \frac{d}{dx}+ eA)^2\psi(x)
\]
where $A(x)$ is the projection of the vector potential along the
wire at point $x$. Using the gauge transformation:
$\psi(x)=\exp(-i\frac{e}{\hbar}\int_{0}^{x}A(x')dx')\chi(x)$,
$\chi(x)$ obeys the Schr\"odinger equation for a free particle,
subjected to the boundary condition:
\mbox{$\chi(l)=\exp(i2\pi \Phi/\Phi_{0})\chi(0)$}, where
\mbox{$\Phi=\frac{e}{\hbar}\int_{0}^{l}A(x')dx'$} is the
magnetic flux through the loop and $\Phi_{0}=h/e$ is the flux quantum.
The single particle energy spectrum is then: 
$E_n=\frac{2\pi^{2}\hbar^{2}}{ml^{2}}(n+\Phi/\Phi_{0})^{2}$, where $n$
is any integer. For generic values of $\Phi/\Phi_{0}$, all energy eigenvalues
are non-degenerate, but all levels are doubly degenerate when $2\Phi/\Phi_{0}$
is an integer (with the exception of the ground-state for integer
$\Phi/\Phi_{0}$). For integer  $\Phi/\Phi_{0}$, the flux through the loop
may be eliminated by a gauge transformation, and the degeneracy is clearly attributed
to the reflection symmetry, which changes the wave-number of an eigenstate
into its opposite. In the case of half-integer $\Phi/\Phi_{0}$,
this geometrical symmetry has to be combined with a gauge transformation
so that $\psi(x)$ is changed into \mbox{$(T\psi)(x)\equiv \psi'(x)=\exp(if(x))\psi(-x)$},
with \mbox{$f(x)=\frac{e}{\hbar}\left(\int_{0}^{-x}A(x')dx'-
\int_{0}^{x}A(x')dx'\right)$}. This transformation
sends an eigenstate with a persistent current into an eigenstate with
the reversed current. We also notice that $\psi'(0)=\psi(0)$
whereas $\psi'(l/2)=-\psi(l/2)$. This has some important consequences
if we wish to generalize these reflection operations to the
elementary loops connecting nodes in $\mathbb Z_{2}$ networks.

\begin{figure}[h]
\includegraphics[width=2.7in]{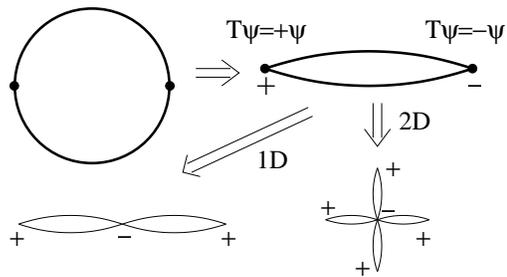}
\caption{This picture shows different realizations of a local $\mathbb Z_2$ symmetry, successively
for a single loop (upper part), and then around sites present in a $1D$ geometry (chain of loops,
lower left) or a $2D$ geometry (square lattice of nodes connected by loops, lower right). 
Signs plus or minus keep track of the relative sign between  
initial and transformed wavefunctions $\psi$ and $T\psi\equiv \psi'$.}
\label{z_2symmetry}
\end{figure}

Let us first consider a one dimensional chain of loops, as shown on the upper part of Fig.~\ref{Z2network}. 
Any link of this chain can be represented by a simple line with plus and minus signs on its extremities,
see Fig.~\ref{z_2symmetry}, reminding us the sign of the transformed wavefunction. 
Suppose now we try to construct a symmetry associated to the reflection for just one loop. 
The simplest way is to combine as before reflection and gauge transformation, but we encounter a major problem: 
this transformation will be no longer local, because we are obliged to change globally the sign of 
the part of the wave-function lying on one side of the transformed loop. 
For an open chain, this non local aspect of the transformation attached to a single loop is
not too disturbing, since it modifies physical observables such as electronic current only locally.
But in the case of periodic boundary conditions, it has the consequence that such a transformation
simply becomes impossible.    
In fact, one can easily avoid this difficulty by considering two transformations on adjacent loops,
see Fig.~\ref{z_2symmetry}. The generalization for any $\mathbb Z_2$ lattice is clear, 
we need to join together negative extremities of all the loops arriving at a given node. 
Because of these phase-factors arising from gauge transformations, the local
symmetry generators are therefore associated to lattice {\em sites} 
and not to the {\em links} as one would infer from purely geometrical considerations.
Note that in lattice gauge theories, local generators of gauge transformations
are also attached to lattice sites, which establishes a close connection between
$\mathbb Z_2$ networks and lattice gauge theories based on the $\mathbb Z_2$ group. 
This viewpoint has been emphasized in the context of Josephson junction arrays~\cite{Doucot03a,Doucot03b}.

In case of an \textit{integer} number of flux quanta per loop ($\Phi=n\Phi_0$) , we have now
$\psi^\prime(0)=\psi(0)$ and $\psi^\prime(l/2)=\psi(l/2)$, so that each link
carries plus signs on both ends. In this case, it is possible to construct local
$\mathbb Z_2$ symmetry generators associated to lattice {\em links} instead of nodes. 
But then, because the single particle Hamiltonian has mostly a dispersive spectrum, 
simultaneous eigenstates of the Hamiltonian and these local generators are extended,
instead of being localized as in the half-flux case.

 \subsection{Constraints on scattering matrices at the nodes} 

\begin{figure}[h]
\includegraphics[width=2.7in]{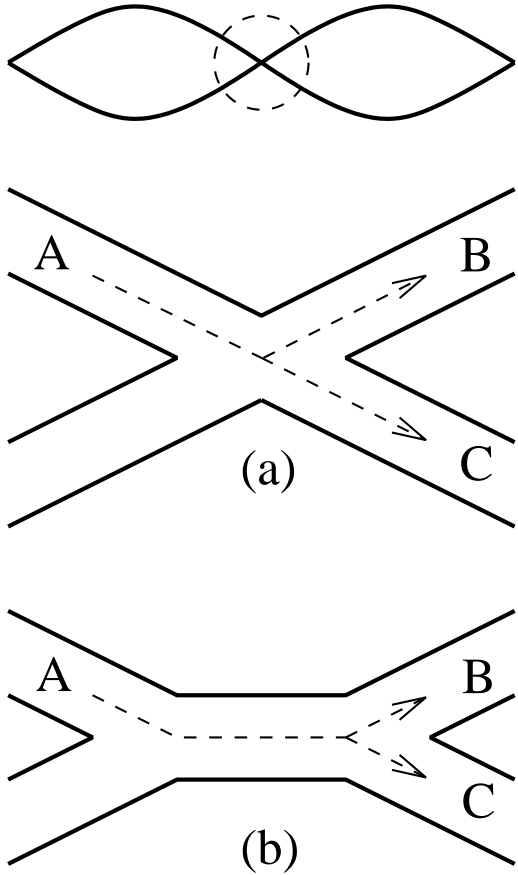}
\caption{This picture illustrates the requirements on the scattering matrix
at a node present in a chain of loops (upper part), in order to impose
a local $\mathbb Z_2$ symmetry. The most direct design (a), is likely
to be incompatible with these constraints, since intuitively, the amplitude 
for an electron to go from A to B is different from the amplitude to go from
A to C. In a slightly modified design (b), these amplitudes are more
likely to be equal, and this is more appropriate for an
experimental implementation of the local $\mathbb Z_2$ symmetry.} 
\label{4node}
\end{figure}

Let us now describe more precisely the local transformation attached to a
given site $i$ on a $\mathbb Z_2$ network. This site belongs to
$n$ elementary loops, so it is the intersection of $2n$ links. 
Since these links are naturally paired, we shall label them by a compound
index $(ij\alpha)$, where $j$ refers to any of the $n$ nearest neighbors of site $i$,
and $\alpha \in \{1,2\}$. On each of these $2n$ links, we choose the origin of
an $x$ coordinate at site $i$, and the projection of the local vector potential
along this link is denoted by $A_{ij\alpha}(x)$. As already discussed, it is
convenient to write the single-particle wave-function as:
\[
\psi_{ij\alpha}(x)=\exp\left(-i\frac{e}{\hbar}\int_{0}^{x}A_{ij\alpha}(x')dx'\right)\chi_{ij\alpha}(x)
\]
along the link $(ij\alpha)$, where an eigenstate with energy $E=\frac{\hbar^2k^2}{2m}$
has the form:
\[
\chi_{ij\alpha}(x)=C_{ij\alpha}e^{-ikx}+D_{ij\alpha}e^{ikx}
\] 
Here, $C_{ij\alpha}$ (resp. $D_{ij\alpha}$) is the amplitude of an incoming (resp. outgoing)
wave with respect to site $i$. 
Since $\chi_{ij\alpha}(x)$ is locally the wave-function of a free particle
with no magnetic field, we may define in a gauge-invariant way a scattering matrix at site $i$ by:
\begin{equation}
D_{ij\alpha}=\sideset{}{^{(i)}}\sum_{j'\alpha'}S^{(i)}_{j\alpha,j'\alpha'}C_{ij'\alpha'}
\end{equation}
which expresses outgoing amplitudes as a function of incoming ones.
Using the notation $\bar 1=2$ and $\bar 2=1$, the local $\mathbb Z_2$ generators
commutes at any site with the single particle Hamiltonian if and only if:
\begin{eqnarray}
S^{(i)}_{j\bar{\alpha},j'\alpha'} & = & S^{(i)}_{j\alpha,j'\alpha'}=S^{(i)}_{j\alpha,j'\bar{\alpha}'}\;\;\; (j \neq j')
\label{firstcond}\\
S^{(i)}_{j\bar{\alpha},j\bar{\alpha}'} & = & S^{(i)}_{j\alpha,j\alpha'}\label{secondcond}
\end{eqnarray}
where $i$ denotes any site, and $j$, $j'$ are arbitrary nearest neighbors of $i$. 
In other words, there should be a perfect symmetry between paired links $ij\alpha$ and $ij\bar{\alpha}$.
The proof of this rather intuitive statement will be given in the next paragraph. Note that in real systems,
nodes always have a finite area, and this constraint may be difficult to implement. For instance,
as shown on Fig.~\ref{4node}, in a chain of loops, it seems very likely that the simplest design
for a node does not satisfy this symmetry. A suggestion for improving this local geometry is
sketched on the bottom of Fig.~\ref{4node}. Generalizations to multichannel quantum
wires are possible, if the underlying inversion symmetry is also performed on the transverse
channels. But we have to restrict ourselves to a regime of weak magnetic fields so that we may neglect
the field effects on orbital motion within a wire.  

Let us now check that conditions~(\ref{firstcond}),(\ref{secondcond}) are all we need.
The transformation attached to site $i$ now reads:
\begin{eqnarray}\label{deftransfi}
\lefteqn{\psi'_{ij\alpha}(x)= } \\
& & \mbox{}-\exp\left(i\sigma_{\alpha}\frac{e}{\hbar}\int_{0}^{x}(A_{ij2}(x')-A_{ij1}(x'))dx'\right)\psi_{ij\bar{\alpha}}(x)
\nonumber
\end{eqnarray}
where $\sigma_{1}=1$ and $\sigma_{2}=-1$.
Using the locally gauge-invariant description in terms of $\chi_{ij\alpha}(x)$, this translates into:
\mbox{$\chi'_{ij\alpha}(x)=-\chi_{ij\bar{\alpha}}(x)$}, which is equivalent to
\mbox{$C'_{ij\alpha}=-C_{ij\bar{\alpha}}$}, and \mbox{$D'_{ij\alpha}=-D_{ij\bar{\alpha}}$}.
This operation leaves the $S$ matrix at site $i$ invariant if and only if:
\begin{equation}
S^{(i)}_{j\bar{\alpha},j'\bar{\alpha}'} = S^{(i)}_{j\alpha,j'\alpha'}
\label{thirdcond}
\end{equation}
Note that this is valid for any $j$ and $j'$ which are nearest neighbors of $i$.
When $j=j'$, we recover condition~(\ref{secondcond}), and when $j \neq j'$,
it is implied by condition~(\ref{firstcond}).
But we should also check that the transformation attached to site $i$ also 
preserves the $S$ matrix at all nearest-neighbor sites $j$ of $i$. Along the link $ij\alpha$,
we have to perform a change of origin, permuting $i$ and $j$, and
replacing $x$ by $y=l_{ij}-x$, where $l_{ij}$ is the common length of
the two links $(ij1)$ and $(ij2)$. Using the fact that the flux through
the loop enclosed by these two links is half a flux quantum modulo $\Phi_0$,
Eq.~(\ref{deftransfi}) becomes:
\begin{eqnarray}\label{deftransfj}  
\lefteqn{\psi'_{ji\alpha}(y)=} \\
& & \exp\left(i\sigma_{\alpha}\frac{e}{\hbar}\int_{0}^{y}(A_{ji2}(y')-A_{ji1}(y'))dy'\right)\psi_{ji\bar{\alpha}}(y)
\nonumber
\end{eqnarray}
which implies \mbox{$\chi'_{ji\alpha}(x)=\chi_{ji\bar{\alpha}}(x)$}, or:
\mbox{$C'_{ji\alpha}=C_{ji\bar{\alpha}}$}, and \mbox{$D'_{ji\alpha}=D_{ji\bar{\alpha}}$}.
This yields \mbox{$S^{(j)}_{i\bar{\alpha},i\bar{\alpha}'} = S^{(j)}_{i\alpha,i\alpha'}$},
that is Eq.~(\ref{secondcond}) after a permutation of $i$ and $j$.
Now, for any site $j'$, nearest neighbor of $j$ and distinct from $i$, the transformation
based at site $i$ does not affect the pair of links $(jj'\alpha)$, so
\mbox{$C'_{jj'\alpha}=C_{jj'\alpha}$}, and \mbox{$D'_{jj'\alpha}=D_{jj'\alpha}$}, which implies
\mbox{$S^{(j)}_{i\bar{\alpha},j'\alpha'}=S^{(j)}_{i\alpha,j'\alpha'}$}
and \mbox{$S^{(j)}_{j'\alpha',i\bar{\alpha}}=S^{(j)}_{j'\alpha',i\alpha}$},
that is Eq.~(\ref{firstcond}) after a permutation of $i$ and $j$.

 \subsection{Irreducible representations of local $\mathbb Z_2$ symmetry}

Let us denote by $U_i$ the unitary transformation attached at site $i$, defined
by Eq.~(\ref{deftransfi}). Clearly, \mbox{$U_i^{2}=\mathbf 1$}, so the eigenvalues of $U_i$ are
$\pm 1$. Furthermore, it is simple to check that $U_i$ and $U_j$ commute.
The only non-trivial case arises when $i$ and $j$ are nearest neighbors. In fact,
comparing Eqs.~(\ref{deftransfi}) and (\ref{deftransfj}) we see that the actions of $U_i$ and $U_j$
on the loop between $i$ and $j$ differ only by a sign. The mutual commutation of these
operators ensures that they can be simultaneously diagonalized, so that
\mbox{$U_i|\Psi\rangle=\epsilon_i|\Psi\rangle$}, with $\epsilon_i=\pm 1$. For single particle
states, the possible collections of eigenvalues $\{\epsilon_i\}$ are very limited. To identify them,
we need to introduce the notion of an Aharonov-Bohm cage. For any lattice site $i$, the cage 
attached to $i$ is composed of all the elementary loops which contain $i$. We have now the following properties:\\
{\bf Property (a):} If $i$ and $j$ are nearest-neighbor sites, and if $U_i|\Psi\rangle=U_j|\Psi\rangle$,
then the wave-function associated to $|\Psi\rangle$ vanishes on the loop joining $i$ and $j$.\\
This is an immediate consequence of the fact that along this loop, $U_i|\Psi\rangle$ and
$U_j|\Psi\rangle$ are described by opposite wave-functions, as already noted (compare Eqs.~(\ref{deftransfi})
and (\ref{deftransfj})).\\
{\bf Property (b):} If $U_i|\Psi\rangle=-|\Psi\rangle$, the wave-function 
$\psi$ attached to $|\Psi\rangle$
vanishes everywhere outside the Aharonov-Bohm cage attached to $i$. \\
This follows directly from the fact that the transformation $U_i$ does not
modify the wave-function outside the Aharonov-Bohm cage attached to $i$.\\ 
{\bf Main Property:} The only possible collections of simultaneous eigenvalues $\{\epsilon_i\}$
are such that exactly one of them is equal to $-1$, all the others being equal to $1$.
If $i$ is the only site with $\epsilon_i=-1$, the corresponding wave-function vanishes everywhere
outside the Aharonov-Bohm cage attached to $i$.\\
Because of property (a), if all $\epsilon_i$'s are equal to $1$, $|\Psi\rangle$ vanishes.
So at least one eigenvalue is negative. Suppose we have two of them: \mbox{$\epsilon_i=\epsilon_j=-1$}
with $i\neq j$. Because of property (b), the wave-function $\psi$ vanishes everywhere
outside the intersection of the cages centered at $i$ and $j$. If this intersection is
empty, then $|\Psi\rangle=0$. If not, then $i$ and $j$ are nearest neighbors. But then, property (a)
implies that $\psi$ vanishes also on the loop joining $i$ and $j$, so $|\Psi\rangle=0$.
The only possibility is therefore to have $\epsilon_i=-1$ and $\epsilon_j=1$ for any
$j\neq i$. From property (b), the wave-function is confined inside the Aharonov-Bohm cage
centered around site $i$. 

When the $S$ matrix of each node satisfies conditions~(\ref{firstcond}) and (\ref{secondcond})
for any wave-vector $k$, we may diagonalize the single-particle Hamiltonian in a localized
basis of Aharonov-Bohm cage wave-functions. Since each of these cages has only a finite
total wire length, the energy spectrum is always discrete. These cages behave as artificial
atoms, at least as long as electrons are not interacting. The interesting feature of such a 
system is that two cages may overlap, which allows for non-trivial interaction effects.

All these results (for $\Phi=\Phi_0/2$) have a simple physical interpretation 
in the light of the Aharonov-Bohm effect. Suppose we have an electron situated at a node $i$. 
If $j$ is a nearest-neighbor of $i$, there are two possible equivalent ways for this electron 
to reach a nearby loop $(j,k)$, using either side of loop $(i,j)$.
But the associated probability amplitudes have opposite signs (as $\exp(i2\pi\Phi/\Phi_0)=-1$) 
and will compensate each other. 
This is why we obtain localized solutions which form a complete basis of the 
Hilbert space for non-interacting electrons.

In case of integer number of flux quanta per loop, the functions within a given irreducible representation 
are not necessarily local, so the previous conclusions do not apply.
It happens that some localized solutions still exist, but they form only  ``half'' of a complete basis of states. 
This analysis could be made using the Kottos and Smilansky formalism~\cite{Kottos97}, but
to save space and since we are mostly interested in half-flux quantum per loop,
we will not give further detail on the integer flux case in this paper. 
 

\section{Interaction effects on Aharonov-Bohm cages}
\label{interactioneffects}

 \subsection{Hamiltonian for interacting electrons in an Aharonov-Bohm cage basis}

In the case of interacting electrons, this simple physical picture has to be deeply modified, because the two sides
of a given loop are no longer equivalent if an electron ``sees'' on one of them another electron and interacts with it. 
So the probability for one electron to hop from an Aharonov-Bohm cage to a nearby one no longer vanishes.
However, this is an interaction-induced effective hopping, and it is by no means obvious that it could lead
to a coherent metallic system in the usual sense. By contrast to lattices such as the dice lattice which
also exhibit the Aharonov-Bohm cage phenomenon, we will now show that on $\mathbb Z_2$ lattices, and
for a large class of electron-electron interactions, the local $\mathbb Z_2$ symmetry is still present.
This puts some rather severe constraints on the amount of delocalization that interactions may induce
for a finite density of electrons. Because of the local $\mathbb Z_2$ symmetry, only {\em pairs}
of electrons are able to hop from one cage to an adjacent one. So on intuitive grounds, a true conducting state
is expected only for attractive interactions, where it is also a superconductor. For repulsive interactions,
we shall see that the system remains an incoherent metal. 

\begin{figure}[h]
\includegraphics[width=2.7in]{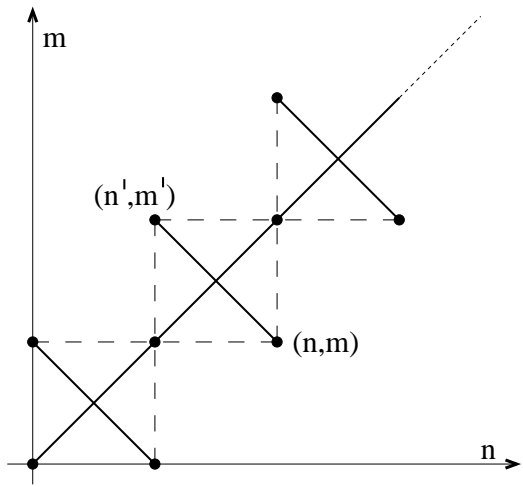}
\caption{In the case of Hubbard like interacting electrons, the only possible cage transitions preserving 
local $\mathbb Z_2$ symmetry can be plotted on the following graph (thin lines), 
where $(n,m)$ and $(n^\prime,m^\prime)$ 
indicate initial and final cage positions, corresponding to a term 
$c^\dagger_{n^\prime}c^\dagger_{m^\prime}c_{m}c_{n}$. 
Dashed lines refer to possible non zero amplitude transitions, compatible with the local nature
of two-electron interaction, but which are forbidden since they violate electron number parity conservation 
resulting from the local $\mathbb Z_2$ symmetry.}
\label{transitions}
\end{figure}

For the sake of simplicity, we shall consider here a local Hubbard repulsive interaction
between electrons. Generalization to a large class of interaction potentials is given
in Section~\ref{superconducting} below. This interaction may be written as:
\begin{equation}
H_{\mathrm{int}}=U\sum_{ij\alpha}\int_{0}^{l_{ij}}\rmd x \:\cpsi_{ij\alpha\uparrow}(x)\cpsi_{ij\alpha\downarrow}(x) 
\apsi_{ij\alpha\downarrow}(x)\apsi_{ij\alpha\uparrow}(x)
\end{equation}
Here $\psi^\dagger_{ij\alpha\uparrow}(x)$ is the second quantization electron creation operator at point $x$ on the wire 
$\alpha=\{1,2\}$ joining nodes $i$ and $j$, with spin projection $\uparrow$ along the quantization axis. 
Let us now introduce operators of creation and annihilation of an electron in 
a given Aharonov-Bohm cage state: $\cc_{i\tau}(x)$ and $\ac_{i\tau}(x)$ respectively,
$\tau$ referring to the spin projection. 
They satisfy the following relations:
\begin{eqnarray}
U_ic_{i\tau}^{\dagger}(x)U_i & = & -c_{i\tau}^{\dagger}(x) \\
U_jc_{i\tau}^{\dagger}(x)U_j & = & c_{i\tau}^{\dagger}(x)\;\;\;i\neq j
\end{eqnarray}
which show that these operators $c_{i\tau}^{\dagger}(x)$ and $\ac_{i\tau}(x)$ carry
a non-trivial local $\mathbb Z_2$ charge.
The precise correspondence between the initial operators $\psi^\dagger_{ij\alpha\uparrow}(x)$ and
the cage operators $c_{i\tau}^{\dagger}(x)$
is as follows:
\begin{equation}
\psi^\dagger_{ij\alpha\tau}(x)=\frac{1}{\sqrt{2}}\left(e^{-i\theta_{ij\alpha}(x)}c_{i\tau}^{\dagger}(x)
+e^{-i\theta_{ji\alpha}(x)}c_{j\tau}^{\dagger}(x)\right)
\label{defcagebasis}
\end{equation}
where \mbox{$\theta_{ij\alpha}(x)=\frac{e}{\hbar}\int_{i}^{x}\mathbf{A}_{ij\alpha}.\mathbf{dr}$}
is a line integral oriented from node $i$ to node $j$.
The local Hubbard interaction in cage basis takes the following form:
\begin{eqnarray}\label{hubbard}
\lefteqn{\psi^\dagger_{ij\alpha\uparrow}(x)\psi^\dagger_{ij\alpha\downarrow}(x) 
\psi_{ij\alpha\downarrow}(x)\psi_{ij\alpha\uparrow}(x)
=} \\
& & \cc_{i\uparrow}(x)\cc_{i\downarrow}(x)\ac_{i\downarrow}(x)\ac_{i\uparrow}(x)+
\cc_{j\uparrow}(x)\cc_{j\downarrow}(x)\ac_{j\downarrow}(x)\ac_{j\uparrow}(x)+\;\:[\mathrm{a}]\nonumber \\
& & \cc_{i\uparrow}(x)\cc_{j\downarrow}(x)\ac_{j\downarrow}(x)\ac_{i\uparrow}(x)+
\cc_{j\uparrow}(x)\cc_{i\downarrow}(x)\ac_{i\downarrow}(x)\ac_{j\uparrow}(x)+\;\:[\mathrm{b}]\nonumber \\
& & \cc_{i\uparrow}(x)\cc_{j\downarrow}(x)\ac_{i\downarrow}(x)\ac_{j\uparrow}(x)+
\cc_{j\uparrow}(x)\cc_{i\downarrow}(x)\ac_{j\downarrow}(x)\ac_{i\uparrow}(x)+\;\:[\mathrm{c}]\nonumber \\
& & \cc_{i\uparrow}(x)\cc_{i\downarrow}(x)\ac_{j\downarrow}(x)\ac_{j\uparrow}(x)+
\cc_{j\uparrow}(x)\cc_{j\downarrow}(x)\ac_{i\downarrow}(x)\ac_{i\uparrow}(x)\;\;\;\;\;\:[\mathrm{d}]\nonumber
\end{eqnarray}
Lines on the right-hand side represent respectively:
an intra-cage Hubbard interaction [a],
an inter-cage Hubbard interaction [b],
a neighbor cage exchange process [c],
a pair propagation process [d].\\
In the case of a one-dimensional chain of loops, these various processes may be visualized
directly on the graph shown on Fig.~\ref{transitions}.

The main feature appearing in this new basis is that the parity of the total 
electron number in {\em each} Aharonov-Bohm cage is {\em separately} conserved.
Because an electron in state  $c_{i\tau}^{\dagger}(x)|0\rangle$ carries a non-trivial
$\mathbb Z_2$ charge, but a pair $c_{i\uparrow}^{\dagger}(x)c_{i\downarrow}^{\dagger}(x)|0\rangle$
belongs to the identity representation of the local $\mathbb Z_2$ group, 
the previous statement simply shows that the local Hubbard interaction commutes
with the local $\mathbb Z_2$ symmetry. In Section~\ref{superconducting}, we show that any
pair potential interaction of the form $V(\mathbf{r},\mathbf{r}')$ is compatible with the
local $\mathbb Z_2$ symmetry provided $V(\mathbf{r},\mathbf{r}')$ does not change
when $\mathbf{r}$ or $\mathbf{r}'$ is replaced by its mirror image with respect to the axis
joining its two nearest nodes. In a previous work, a similar statement was reached for a
rather special case, namely a chain of rhombi in a tight-binding description~\cite{Doucot02}.
The present formulation is much more general.

What are the consequences of this local symmetry on an interacting system?
As in~\cite{Vidal00,Vidal01}, let us first consider a pair of electrons, with the local 
Hubbard interaction. We have three possible situations. First, the two electrons
may be located in two cages $i$ and $j$ which do not share common links, that is
$i$ and $j$ are not nearest neighbors. Then they do not ``see'' each other
and remain localized in their respective cages. A second possibility occurs
when $i$ and $j$ are distinct but nearest neighbors. As term [c] in Eq.~(\ref{hubbard})
above shows, the two electrons can exchange their cages, but such state remains
localized under time evolution. Finally, when $i=j$, term [d] in Eq.~(\ref{hubbard})
indicates that a delocalized two electron state is now possible in spite of the
purely localized nature of the single particle spectrum and the repulsive character 
of the interaction.     

Of course, the most interesting question is to consider a system with a finite
electronic density. As we have seen, because of the local $\mathbb Z_2$ symmetry,
only pairs of electrons can propagate in a $\mathbb Z_2$ network.
In particular, the single electron propagator \mbox{$-i\langle \ac_{i\tau}(x,t)\cc_{j\tau}(x',t')\rangle$}
vanishes when $i\neq j$, since it is not $\mathbb Z_2$ invariant, and it is very difficult
to break spontaneously a local symmetry~\cite{Elitzur75}.
This shows that the low-energy behavior of an electron fluid on a $\mathbb Z_2$ network
is {\em not} described by the Landau model of a Fermi liquid. As we shall see in
section~\ref{RG} below, it corresponds most likely to a new form of incoherent metal.
But at this stage, the most natural thing to check is whether a superconducting instability
(induced by term [d] in Eq.~(\ref{hubbard})) occurs or not. 

 \subsection{Mean-field analysis of superconducting instability}
 \label{superconducting}

To begin this discussion, let us first consider a larger class of electron-electron
interaction. The most general two-body potential interaction respecting the global 
spin rotation symmetry reads:
\begin{eqnarray}
\lefteqn{\hat{V}=\sum_{ij,i'j'}\sum_{\alpha\alpha'}\sum_{\tau\tau'}\int_{0}^{l_{ij}}\rmd x \int_{0}^{l'_{ij}}\rmd x' \;
V_{ij,i'j'}^{\alpha\alpha'}(x,x')} \nonumber \\
& & \cpsi_{ij\alpha\uparrow}(x)\cpsi_{i'j'\alpha'\downarrow}(x') 
\apsi_{i'j'\alpha'\downarrow}(x')\apsi_{ij\alpha\uparrow}(x)
\label{generalinteraction}
\end{eqnarray}
In this sum, $(ij)$ and $(i'j')$ are two pairs of nearest-neighbor nodes. As before,
$\alpha$ (resp. $\alpha'$) labels one of the two links composing the loop $(ij)$
(resp. $(i'j')$).
Let us use the cage basis, defined in Eq.~(\ref{defcagebasis}) to rewrite this interaction.
For this purpose, the first step is to express the local electronic density as:
\begin{eqnarray*}
\lefteqn{\cpsi_{ij\alpha\tau}(x)\apsi_{ij\alpha\tau}(x) = \frac{1}{2}
(\cc_{i\tau}(x)\ac_{i\tau}(x)+\cc_{j\tau}(x)\ac_{j\tau}(x))} \\
& & \mbox{} + \frac{1}{2}(e^{-i\theta_{ij\alpha}}\cc_{i\tau}(x)\ac_{j\tau}(x)+
e^{-i\theta_{ji\alpha}}\cc_{j\tau}(x)\ac_{i\tau}(x))
\end{eqnarray*} 
Because the flux through each elementary loop is half a flux quantum,
\mbox{$e^{-i\theta_{ij\alpha}}=-e^{-i\theta_{ij\bar{\alpha}}}$}.
The interaction~(\ref{generalinteraction}) preserves the local $\mathbb Z_{2}$
symmetry if and only if the following conditions are satisfied:
\begin{equation}
V_{ij,i'j'}^{\alpha\alpha'}(x,x')=V_{ij,i'j'}^{\bar{\alpha}\alpha'}(x,x')=V_{ij,i'j'}^{\alpha\bar{\alpha}'}(x,x')
\label{firstcondint}
\end{equation}
provided the pairs $(ij)$ and $(i'j')$ are distinct, and:
\begin{equation}
V_{ij,ij}^{\alpha\alpha'}(x,x')=V_{ij,ij}^{\bar{\alpha}\bar{\alpha}'}(x,x')
\label{secondcondint}
\end{equation}
The first condition implies that 
\[\sum_{\alpha}e^{-i\theta_{ij\alpha}}V_{ij,i'j'}^{\alpha\alpha'}(x,x')=0\]
when $(ij)$ and $(i'j')$ are distinct, and the second condition that
\[\sum_{\alpha,\alpha'}e^{-i\theta_{ij\alpha}}V_{ij,ij}^{\alpha\alpha'}(x,x')=0\]
Therefore, all the terms which do not preserve the parity of the total electron
number in each cage are eliminated. When both conditions~(\ref{firstcondint}), (\ref{secondcondint})
on the interaction potential are satisfied, the interaction term~(\ref{generalinteraction})
now becomes:
\begin{widetext}
\begin{eqnarray}
\hat{V} & = & \sum_{ij,i'j'}\sum_{\alpha\alpha'}\sum_{\tau\tau'}\int_{0}^{l_{ij}}\rmd x \int_{0}^{l'_{ij}}\rmd x' \;
V_{ij,i'j'}^{\alpha\alpha'}(x,x') \nonumber \\
& & \times\Bigg\{:\left[\cc_{i\tau}(x)\ac_{i\tau}(x)+\cc_{j\tau}(x)\ac_{j\tau}(x)\right]
\left[\cc_{i'\tau'}(x')\ac_{i'\tau'}(x')+\cc_{j'\tau'}(x')\ac_{j'\tau'}(x')\right]: \\
& & \mbox{} + \delta_{ij,i'j'} :\left[e^{-\rmi\theta_{ij\alpha}}\cc_{i\tau}(x)\ac_{j\tau}(x)+
e^{-\rmi\theta_{ji\alpha}}\cc_{j\tau}(x)\ac_{i\tau}(x)\right]
\left[e^{-\rmi\theta_{ij\alpha'}}\cc_{i\tau}(x')\ac_{j\tau}(x)+
e^{-\rmi\theta_{ji\alpha'}}\cc_{j\tau}(x)\ac_{i\tau}(x)\right]:\Bigg\} \nonumber
\label{Z2geninteraction}
\end{eqnarray}
\end{widetext}
We have used the standard notation $:\mathcal{O}:$ for the normal 
ordering of the operator $\mathcal{O}$. 

Let us now consider the possibility of a superconducting instability.
In a variational approach, we would use a Hartree-Fock-Bogoliubov
trial state. In the presence of pairing, the average kinetic energy increases,
so it would have to be compensated by a lowering in the average interaction 
energy. As usual, this quantity involves averages of the form
\mbox{$\langle\cc\ac\rangle\langle\cc\ac\rangle$} and of the form
\mbox{$\langle\cc\cc\rangle\langle\ac\ac\rangle$}. In the spirit of
a mean field theory, we shall keep only the latter terms. Once again we also
argue that the local $\mathbb Z_2$ symmetry is not broken, even in the
hypothetical superconducting state. Physically, this is reasonable since the
size of the Aharononv-Bohm cages being finite, there is a finite gap between
the $\mathbb Z_2$ sector with $\epsilon_{i}=1$ and the sector with $\epsilon_{i}=-1$.
As discussed below, this gap can be as large as a few Kelvin degrees for a
GaAs $\mathbb Z_2$ network. With an unbroken $\mathbb Z_2$ symmetry,
the superconducting order parameter $\langle \cc_{i\tau}(x)\cc_{j\tau}(x')\rangle$
vanishes if $i\neq j$.

Let us first consider a singlet pairing. This leads to:
\begin{eqnarray*}
\langle \cc_{i\uparrow}(x)\cc_{i\uparrow}(x')\rangle & = & 
\langle \cc_{i\downarrow}(x)\cc_{i\downarrow}(x')\rangle = 0 \\
\langle \cc_{i\uparrow}(x)\cc_{i\downarrow}(x')\rangle & = & 
-\langle \cc_{i\downarrow}(x)\cc_{i\uparrow}(x')\rangle = \Delta_{i}(x,x')
\end{eqnarray*}
In this case, the local contribution of Cooper pairing to the trial potential
energy is:
\begin{equation}
V_{ij,i'j'}^{\alpha\alpha'}(x,x')|e^{-\frac{i}{2}\theta_{ij\alpha}}\Delta_{i}(x,x')
+e^{-\frac{i}{2}\theta_{ji\alpha}}\Delta_{j}(x,x')|^{2}
\label{localpairenergy}
\end{equation}
In the case of triplet pairing, we have a complex vector $\boldsymbol{\Delta}_{j}(x,x')$
defined by:
\begin{eqnarray*}
\boldsymbol{\Delta}^{x}_{j}(x,x') & = & \frac{1}{\sqrt{2}}\langle-\cc_{j\uparrow}(x)\cc_{j\uparrow}(x')+
\cc_{j\downarrow}(x)\cc_{j\downarrow}(x')\rangle\\ 
\boldsymbol{\Delta}^{y}_{j}(x,x') & = & \frac{i}{\sqrt{2}}\langle\cc_{j\uparrow}(x)\cc_{j\uparrow}(x')+
\cc_{j\downarrow}(x)\cc_{j\downarrow}(x')\rangle\\ 
\boldsymbol{\Delta}^{z}_{j}(x,x') & = & \frac{1}{\sqrt{2}}\langle\cc_{j\uparrow}(x)\cc_{j\downarrow}(x')+
\cc_{j\downarrow}(x)\cc_{j\uparrow}(x')\rangle 
\end{eqnarray*}
For triplet pairing, the orbital angular momentum of a Cooper pair is odd, so
$\boldsymbol{\Delta}_{j}(x,x')=-\boldsymbol{\Delta}_{j}(x',x)$. It is simple to check that the above 
expression~(\ref{localpairenergy})  still applies up to replacing the complex scalar
$\Delta_{j}(x,x')$ by the complex vector $\boldsymbol{\Delta}_{j}(x,x')$.

To conclude, we show that the part of the trial energy which is directly
sensitive to a $\mathbb Z_2$ invariant Cooper pairing amplitude is positive
for a positive interaction potential compatible with the local $\mathbb Z_2$
symmetry. So, at least in a mean-field picture, a superconducting instability
cannot occur. This statement was not a priori obvious, given the existence
of delocalized states for the two-particle problem, where both particles
are located in the same cage at any time, even with a repulsive interaction.

 \subsection{Discussion for networks in semiconductor heterostructures}

For applications to real systems, it is important to present some order of
magnitude estimates. For wires etched on a 2D electron gas at a GaAs/GaAlAs
interface, with a typical carrier density of 3 $10^{11}$ $\mathrm{cm}^{-2}$
and a transverse width $w=100$ nm, the number of transverse conduction channels
$N=k_{\mathrm{F}}w/\pi$ is around 4. For a regular network of single channel 
wires, we have investigated interaction effects on the low-energy scattering matrix
at the nodes~\cite{Kazymyrenko}. The renormalization of the transmission coefficient
$T(\Lambda)$ as a function of a typical energy scale $\Lambda$ is given by:
\begin{equation}
T(\Lambda)=\frac{T_{0}(\Lambda/\Lambda_{0})^{2\alpha}}{R_{0}+T_{0}(\Lambda/\Lambda_{0})^{2\alpha}},
\;\;\;\;(R_{0}+T_{0}=1)
\label{scalingT}
\end{equation}
where $\alpha=\tilde{U}(k\rightarrow 0)/(hv_{\mathrm{F}})$ is proportional
to the Fourier transform of the interaction potential in the long wave-length limit.
This result corresponds to nodes having a complete symmetry under any permutation 
of the incident wires. Note that it has the same form as for a single elastic scatterer
in a Luttinger liquid~\cite{Kane92,Yue94}, or a single crossing connecting several
semi-infinite Luttinger liquids~\cite{Lal02}. For a single wire of width $w$ and length $L$,
\mbox{$\tilde{U}(k\rightarrow 0)\simeq \frac{e^{2}}{4\pi \epsilon_{0}\epsilon_{r}}\ln(2L/w)$}.
Choosing typical values as $m^{*}=0.067$ $m_e$ and $\epsilon_{r}=12.9$ for GaAs,
$L=2$ $\mu$m and $w=100$ nm, we obtain $\alpha=0.4$ which is a rather large value.
This shows that the Coulomb energy is nearly as large as the kinetic energy of
conduction electrons. 

As usual for 1D electron systems with repulsive interactions,
the leading instability occurs because of the divergence of the static 
charge and spin susceptibilities at the wave-vector $q=2k_{\mathrm{F}}$ and favors
the formation of a spin-density wave. A derivation of this mean-field transition on a $\mathbb Z_2$
network is sketched in Appendix~\ref{AppSDW}. But with $\alpha=0.4$,  the critical temperature
for the onset of this mean-field instability is not much smaller than the Fermi energy
(see for instance Eq.~(\ref{meanfieldTc})),
which shows that the system is still in a one-dimensional regime.
Indeed, the cross-over from one to two dimensions takes place when the 1D thermal length
$L_{\mathrm{T}}=hv_{\mathrm{F}}/k_{\mathrm{B}}T$ becomes comparable to the
distance $L$ between nearest-neighbor nodes. In other words, the corresponding
energy scale $k_{\mathrm{B}}T$ should be as low as the energy level splittings
inside Aharonov-Bohm cages, that is of the order of $hv_{\mathrm{F}}/L$.
Since $\lambda_{\mathrm{F}}=50$ nm ($L/\lambda_{\mathrm{F}}=40$), the cross-over to 2D behavior appears at an energy scale
one order of magnitude smaller than the mean-field Peierls instability scale. 
In practice however, because of the small effective mass in GaAs, the low energy scale
$hv_{\mathrm{F}}/L$ can be of the order of a few Kelvin degrees. This scale sets also
the order of magnitude of the $\mathbb Z_2$ gap already mentioned.

It is well-known that in a purely 1D system, fluctuations
in the Peierls and in the Cooper channels have a tendency to neutralize each other,
and the result is in general some kind of Luttinger liquid~\cite{Solyom79}. 
As mean-field analysis fails to predict such a behavior,
we should therefore revisit the result of the previous section using an approach
which is able to treat both Peierls and Cooper channels
on an equal footing. This is exactly what the renormalization group (RG) is
able to provide, at least in the weak interaction limit. 
To be completely honest, it is not at all clear that a perturbative RG
can give reliable quantitative estimates when $\alpha$ is as large as 0.4. 
But for the sake of establishing the presence or the absence of an instability,
we expect it to produce a valid qualitative picture, specially in the case, as we
shall see below, where it predicts an instability to a gapped state already for small coupling.
It might happen, as in some Hubbard models with extended internal symmetries~\cite{Boulat},
that some features of the gapped phase are modified when the coupling becomes large
compared to the electronic bandwidth. But in this example, the gap never closes
when the system evolves from weak to strong coupling.


\section{Renormalization group analysis}
\label{RG}

 \subsection{Low-energy model}

A complete RG analysis for our original system would be very complicated,
because of the absence of translation invariance inside Aharonov-Bohm cages.
In particular, network nodes induce Friedel density oscillations in the
electron fluid, which renormalize the associated single electron scattering matrices
in the presence of interactions. This mechanism is precisely at the origin 
of the scaling shown in Eq.~(\ref{scalingT})~\cite{Yue94,Lal02,Kazymyrenko}.
Physically, this interaction effect shows that wires have a tendency to become disconnected
at low energies, which stops below the cross-over into the 2D regime. 
As shown previously~\cite{Kazymyrenko}, an electronic wire network becomes 
an insulator for commensurate filling factors, or simply highly resistive, 
with a weakly dispersive band crossing the Fermi level for generic
filling factors. 
However, for a system with local $\mathbb Z_2$ symmetry, this renormalization
mostly modifies the internal dynamics inside a given Aharonov-Bohm cage, and not
the general structure of the electron-electron interaction as it appears in Eq.~\ref{hubbard}.
At this point, we shall now assume that details of this internal dynamics inside
Aharonov-Bohm cages do not play much role in the analysis of interaction effects
at low energy, provided the local $\mathbb Z_2$ symmetry manifested in 
Eq.~(\ref{hubbard}) is still respected.  
The validity of this assumption will be examined in more detail at the end
of this section, once the results of the RG treatment are available.

This leads us to consider the following model, where each Aharonov-Bohm cage is
replaced by an infinite 1D Luttinger liquid, described by the electronic fields
$\cc_{i\tau}(x)$, $\ac_{i\tau}(x)$, $i$ being a cage label and $\tau$ the spin
projection. The internal energy levels of this collection of modified Aharonov-Bohm
cages is naturally described by the following kinetic Hamiltonian:
\begin{widetext}
  \begin{equation}
    \label{eq:low_energy_H0}
    H_0 = \int_{-\infty}^\infty \rmd x \sum_j \sum_\tau 
    :\Bigg[ 
    \cc_{\rmr,j,\tau}(x)\frac{1}{\rmi}\frac{\rmd}{\rmd x}
    \ac_{\rmr,j,\tau}(x)
    -\cc_{\rml,j,\tau}(x)\frac{1}{\rmi}\frac{\rmd}{\rmd x}
    \ac_{\rml,j,\tau}(x) \Bigg]:
  \end{equation}
\end{widetext}
where $hv_{\mathrm{F}}$ has been set equal to unity.
As usual, right and left moving fields have been introduced,
labelled by R and L respectively, so that 
\mbox{$\cc_{i\tau}(x)=\cc_{\mathrm{R},i,\tau}(x)+\cc_{\mathrm{L},i,\tau}(x)$}.
Note that replacing finite Aharonov-Bohm cages by infinite ones
is only valid in the 1D regime, when we can neglect the discrete nature
of the single particle spectrum. The physical motivation for 
doing this is that as discussed in the previous section, the typical 
interaction scale is larger than the $\mathbb Z_2$ gap in present experiments.
On the other hand, the perturbative RG used here requires
small interactions in comparison to the single electron bandwidth.
Because of the large value of $L/\lambda_{\mathbf{F}}$, these two
requirements are mutually compatible.

Two nearby cages $i$ and $j$ are coupled by a two-body interaction
subjected to the following constraints: it has to preserve the local
$\mathbb Z_2$ symmetry or equivalently, to conserve the parity of the total
number of electrons in each cage, and to simplify the treatment, to
preserve also the translational symmetry within cages. Both requirements
are satisfied by the following Hubbard-like interaction:
\begin{widetext}
\begin{eqnarray*}
\lefteqn{H_\mathrm{int}=U \int_{-\infty}^\infty \rmd x \sum_{i,j}} \\
& &:\Big[ 
\cc_{i\uparrow}(x)\cc_{i\downarrow}(x)\ac_{i\downarrow}(x)\ac_{i\uparrow}(x)+
\cc_{j\uparrow}(x)\cc_{j\downarrow}(x)\ac_{j\downarrow}(x)\ac_{j\uparrow}(x)+ 
\cc_{i\uparrow}(x)\cc_{j\downarrow}(x)\ac_{j\downarrow}(x)\ac_{i\uparrow}(x)+
\cc_{j\uparrow}(x)\cc_{i\downarrow}(x)\ac_{i\downarrow}(x)\ac_{j\uparrow}(x) \\\
& & \mbox{}+\cc_{i\uparrow}(x)\cc_{j\downarrow}(x)\ac_{i\downarrow}(x)\ac_{j\uparrow}(x)+
\cc_{j\uparrow}(x)\cc_{i\downarrow}(x)\ac_{j\downarrow}(x)\ac_{i\uparrow}(x)+ 
\cc_{i\uparrow}(x)\cc_{i\downarrow}(x)\ac_{j\downarrow}(x)\ac_{j\uparrow}(x)+
\cc_{j\uparrow}(x)\cc_{j\downarrow}(x)\ac_{i\downarrow}(x)\ac_{i\uparrow}(x) \Big]: 
\end{eqnarray*}
\end{widetext}
Using the decomposition \mbox{$\cc_{i\tau}(x)=\cc_{\mathrm{R},i,\tau}(x)+\cc_{\mathrm{L},i,\tau}(x)$}
generates many terms, among which we shall neglect two subsets. The first subset
contains all the purely forward scattering processes on a given side of the
Fermi-surface, i.e. of the form $\cc_{\mathrm{R}}\cc_{\mathrm{R}}\ac_{\mathrm{R}}\ac_{\mathrm{R}}$ or  
$\cc_{\mathrm{L}}\cc_{\mathrm{L}}\ac_{\mathrm{L}}\ac_{\mathrm{L}}$. In perturbation theory,
they do not give singular terms in the low-energy limit, so we shall ignore them.
The second subset contains all Umklapp terms, of the form 
\mbox{$\cc_{\mathrm{R}}\cc_{\mathrm{R}}\ac_{\mathrm{L}}\ac_{\mathrm{R}}$} 
(resp. \mbox{$\cc_{\mathrm{R}}\cc_{\mathrm{R}}\ac_{\mathrm{L}}\ac_{\mathrm{L}}$}) 
which are relevant only for a filled (resp. half-filled) single electron band.  
The most interesting terms are of the type
\mbox{$\cc_{\mathrm{R}}\cc_{\mathrm{L}}\ac_{\mathrm{L}}\ac_{\mathrm{R}}$} which
induce logarithmic divergences to the dressed two-particle interaction in the low
energy limit. These terms are adequately treated by the renormalization group. 
Introducing now generalized charge and spin densities defined as:
\begin{eqnarray}
  \label{eq:def_rho_S}
  \rho_{\rmr(\rml),n,n'}(x)=\sum_{\tau,\tau'} \cc_{\rmr(\rml),n,\tau}(x) 
  \delta_{\tau,\tau'} \ac_{\rmr(\rml),n',\tau'}(x),\\
  \bS_{\rmr(\rml),n,n'}(x)=\sum_{\tau,\tau'} \cc_{\rmr(\rml),n,\tau}(x) 
  \boldsymbol{\sigma}_{\tau,\tau'} \ac_{\rmr(\rml),n',\tau'}(x),
\end{eqnarray}
where $\boldsymbol{\sigma}=(\sigma^x,\sigma^y,\sigma^z)$ are the usual Pauli
matrices, we may finally write the interaction term in the form:
\begin{widetext}
\begin{eqnarray}
\label{eq:low_energy_Hint}
    H_\mathrm{int} &=& U \int_{-\infty}^\infty \rmd x \sum_n :\Big[ 
    2 \rho_{\rmr,n,n}(x) \rho_{\rml,n,n}(x) 
    - 2\bS_{\rmr,n,n}\bS_{\rml,n,n}\\
    && + \rho_{\rmr,n,n}(x) \rho_{\rml,n+1,n+1}(x) 
    - \bS_{\rmr,n,n}\bS_{\rml,n+1,n+1}
    + \rho_{\rmr,n+1,n+1}(x) \rho_{\rml,n,n}(x) 
    - \bS_{\rmr,n+1,n+1}\bS_{\rml,n,n}\nonumber\\
    && + \rho_{\rmr,n+1,n}(x) \rho_{\rml,n,n+1}(x) 
    - \bS_{\rmr,n+1,n}\bS_{\rml,n,n+1}
    + \rho_{\rmr,n,n+1}(x) \rho_{\rml,n+1,n}(x) 
    - \bS_{\rmr,n,n+1}\bS_{\rml,n+1,n}\nonumber\\
    && + \rho_{\rmr,n+1,n}(x) \rho_{\rml,n+1,n}(x) 
    - \bS_{\rmr,n+1,n}\bS_{\rml,n+1,n}
    + \rho_{\rmr,n,n+1}(x) \rho_{\rml,n,n+1}(x) 
    - \bS_{\rmr,n,n+1}\bS_{\rml,n,n+1} \Big]:.\nonumber
  \end{eqnarray}
\end{widetext}

Here, we have chosen a particular 1D geometry, namely a chain of loops.
Although the RG approach could be applied to any regular lattice
of Luttinger liquid-like cages, we shall from now on focus on this example.
The results we shall present shortly suggest that the large scale lattice geometry
does not affect much the physical properties of such a system. 


\subsection{Renormalization group equations}

\begin{figure*}[t]
  \centering
  \includegraphics[width=7cm]{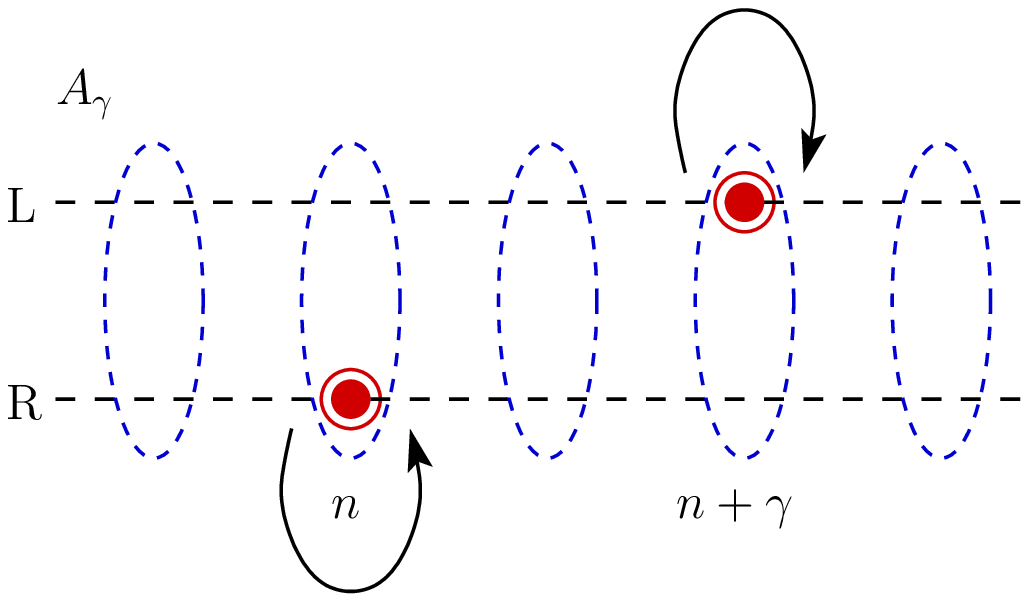}\hspace{1cm}\includegraphics[width=7cm]{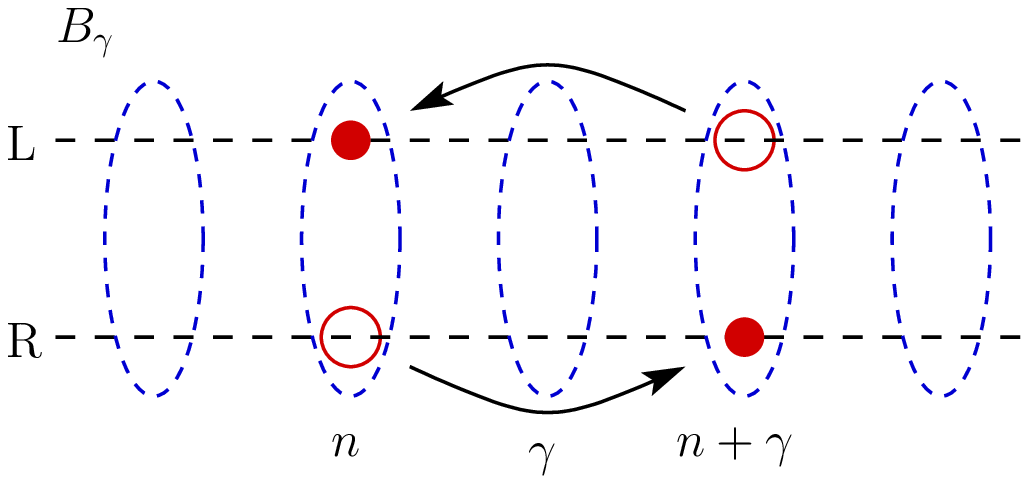}
  \includegraphics[width=7cm]{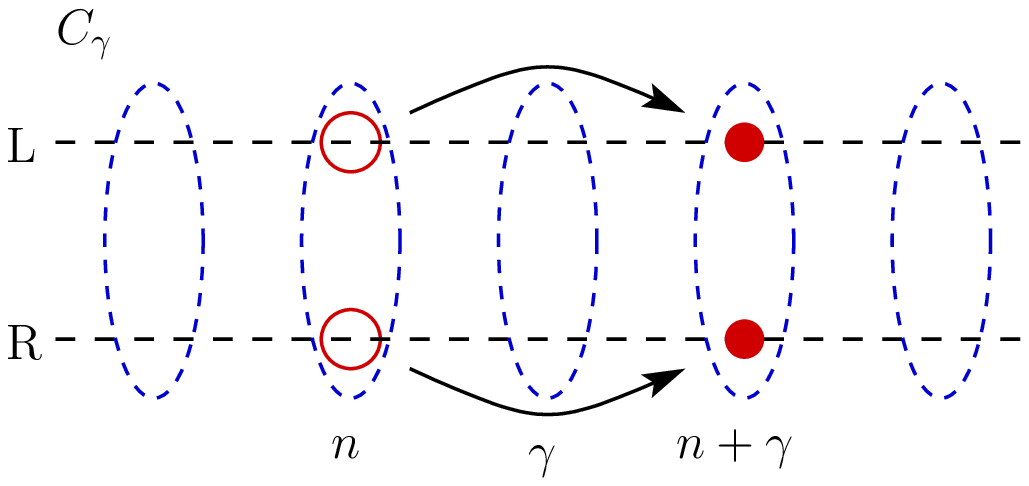}\hspace{1cm}\includegraphics[width=7cm]{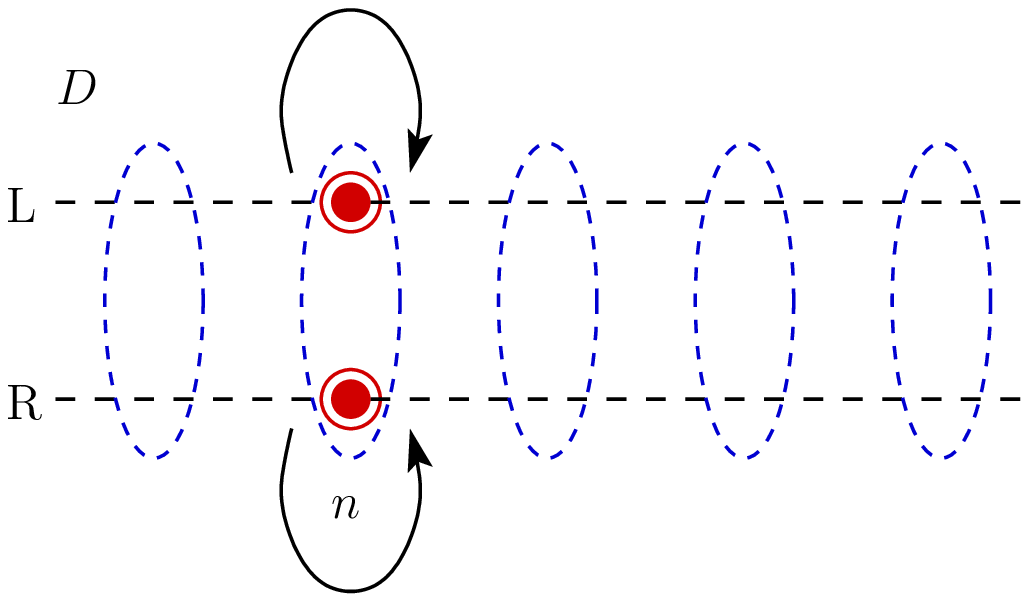}
  \caption{Schematic representation of the four interactions compatible with
    the local $\mathbb{Z}_2$-symmetry, given in Eq.~(\ref{eq:flowing_Hint}).
    Each dashed ellipse represents a cage. 
    An empty (full) circle represents a destruction (creation) operator.
    Coupling $A$ is an inter-cage interaction, coupling $B$ a cage exchange, 
    coupling $C$ a pair hopping, and coupling $D$ an intra-cage interaction.}
  \label{fig:interactions}
\end{figure*}

We shall now briefly derive one-loop RG equations. We refer the reader to 
Ref.~\onlinecite{Dusuel02} and references therein for general details about 
the RG technique. Here we focus on what happens in the space of cage positions,
since this is what makes the originality of our work.
The microscopic Hamiltonian 
(\ref{eq:low_energy_H0}-\ref{eq:low_energy_Hint}) serves as the 
high energy initial condition of RG flows. The kinetic part of the Hamiltonian 
is unrenormalized at one-loop order, while the interaction part flows.
During the renormalization process, the interactions will develop a 
non-trivial dependence in the cage positions. The local 
$\mathbb{Z}_2$-symmetry as well as the translational symmetry of the initial
Hamiltonian (\ref{eq:low_energy_H0}-\ref{eq:low_energy_Hint}) will remain
symmetries of the renormalized Hamiltonian. 
As a consequence, the flowing Hamiltonian is constrained to be of the 
following form
\begin{widetext}
  \begin{eqnarray}
    \label{eq:flowing_H0}
    H_0(\Lambda) &=& \int_{-\Lambda}^\Lambda \frac{\rmd k}{2\pi} 
    \sum_n \sum_\tau 
    :\Big[k \cc_{\rmr,n,\tau}(k)\ac_{\rmr,n,\tau}(k)
    -k \cc_{\rml,n,\tau}(k)\ac_{\rml,n,\tau}(k) \Big]:\\
    \label{eq:flowing_Hint}
    H_\mathrm{int}(\Lambda) &=& \int_{-\Lambda}^\Lambda \frac{\rmd q}{2\pi} 
    \sum_{n}: \Bigg\{ \sum_{\gamma\neq 0}\Big[ 
    A^\rmc_\gamma(\Lambda)\rho_{\rmr,n,n}(q)\rho_{\rml,n+\gamma,n+\gamma}(-q)
    + A^\rms_\gamma(\Lambda)\bS_{\rmr,n,n}(q)\bS_{\rml,n+\gamma,n+\gamma}(-q)
    \\
    && 
    + B^\rmc_\gamma(\Lambda)\rho_{\rmr,n+\gamma,n}(q)\rho_{\rml,n,n+\gamma}(-q)
    + B^\rms_\gamma(\Lambda)\bS_{\rmr,n+\gamma,n}(q)\bS_{\rml,n,n+\gamma}(-q)
    \nonumber\\
    &&
    + C^\rmc_\gamma(\Lambda)\rho_{\rmr,n+\gamma,n}(q)\rho_{\rml,n+\gamma,n}(-q)
    + C^\rms_\gamma(\Lambda)\bS_{\rmr,n+\gamma,n}(q)\bS_{\rml,n+\gamma,n}(-q)
    \Big]\nonumber\\
    && + D^\rmc(\Lambda) \rho_{\rmr,n,n}(q)\rho_{\rml,n,n}(-q)
    + D^\rms(\Lambda) \bS_{\rmr,n,n}(q)\bS_{\rml,n,n}(-q)
    \Bigg\}:\nonumber.
  \end{eqnarray}
\end{widetext}
Note that the hermiticity of the Hamiltonian implies that
$C^\rmcs_{-\gamma}=C^\rmcs_\gamma$, and the Right/Left symmetry gives
$A^\rmcs_{-\gamma}=A^\rmcs_\gamma$ and $B^\rmcs_{-\gamma}=B^\rmcs_\gamma$.
The action of the four types of locally $\mathbb{Z}_2$-symmetric couplings 
$A_\gamma$, $B_\gamma$, $C_\gamma$ and $D$ (forgetting about their charge/spin 
nature) are represented in Fig.~\ref{fig:interactions}.

In Eqs.~(\ref{eq:flowing_H0}-\ref{eq:flowing_Hint}) we used the Fourier 
transform of the field operators which is defined by
\begin{equation}
  \ac_{\rmr(\rml),n,\tau}(k)=\int_{-\infty}^\infty \rmd x \exp(-\rmi k x)
  \ac_{\rmr(\rml),n,\tau}(x).
\end{equation}
The Fourier transform of the generalized charge density, for example, 
is given by 
\begin{eqnarray}
  \label{eq:TF_rho}
  &&\rho_{\rmr(\rml),n,n'}(q) = \int_{-\infty}^\infty \rmd x \exp(\rmi q x)
  \rho_{\rmr(\rml),n,n'}(x)\\
  &&\hspace{1cm} = \int_{-\Lambda}^\Lambda \frac{\rmd k}{2\pi} \sum_\tau 
  \cc_{\rmr(\rml),n,\tau}(k+q) \ac_{\rmr(\rml),n',\tau}(k),\nonumber
\end{eqnarray}
All momenta have been restricted to be smaller in absolute value than the 
cut-off $\Lambda$, in order to regularize ultra-violet divergences.
Note that the Fermi momentum has been set to zero, which is allowed since we 
do not consider Umklapp processes.
We have not denoted any momentum dependence of the interactions, 
because in the RG treatment, all momenta are taken at the Fermi surface.

\begin{figure*}[t]
   \centering
  \includegraphics[width=5.5cm]{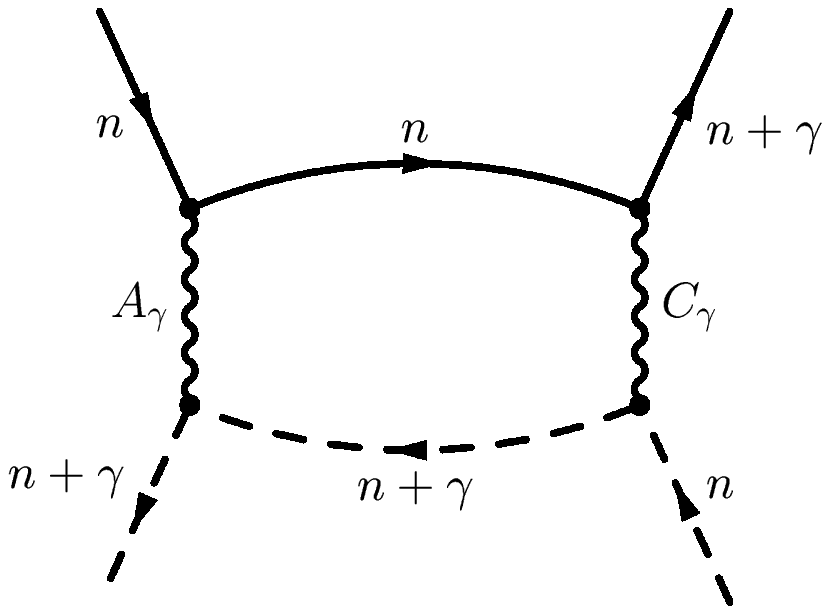}\quad\includegraphics[width=5.5cm]{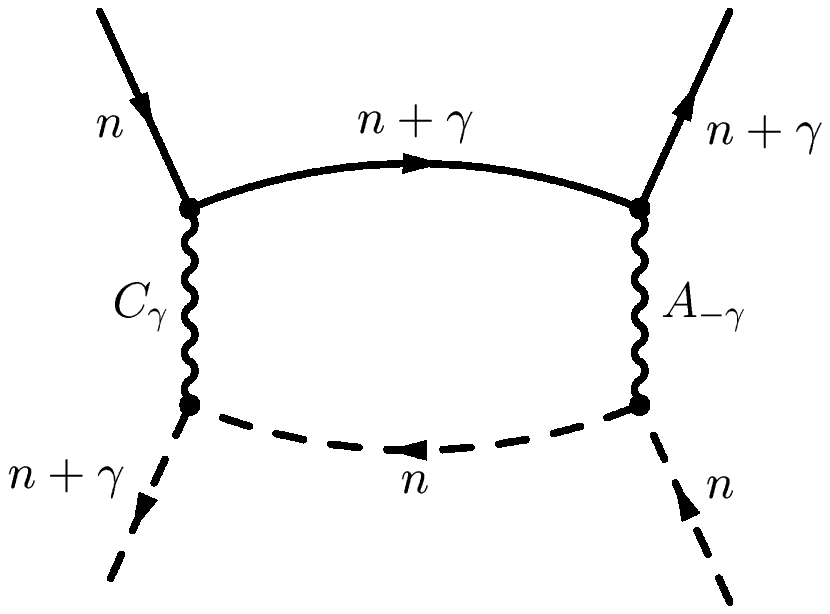}\quad\includegraphics[width=5.5cm]{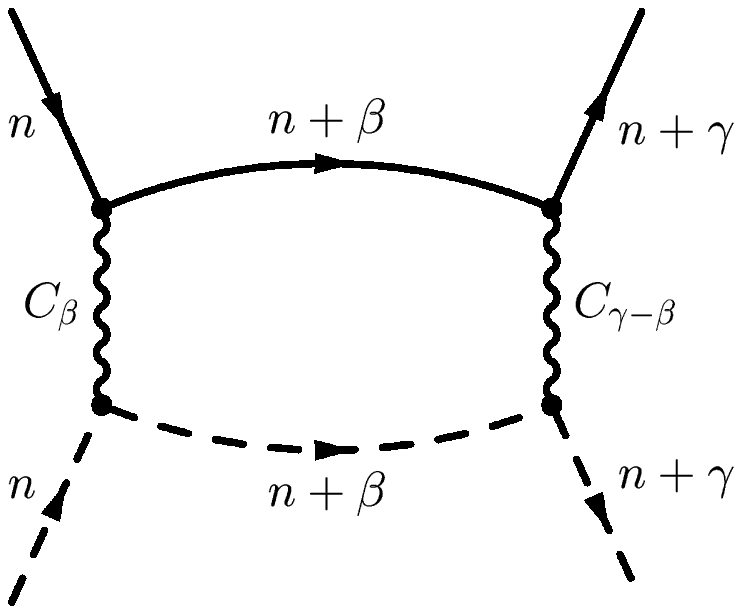}
  \caption{Graphs renormalizing the pair hopping term $C_\gamma$. The 
  left and middle graphs are particle-hole bubbles, and the right one is a
  particle-particle bubble. For the latter, $\beta$ is summed over, and 
  when $\beta=0$ or $\gamma$, the quantity $C_0$ that
  appears should be understood as $D$, so as to yield 
  Eq.~(\ref{eq:RG_equations_C}).}
\label{fig:renC}
\end{figure*}

The high-energy Hamiltonian is defined at scale $\Lambda_0$, and
the initial conditions of the RG flow, at this scale, are found from 
(\ref{eq:low_energy_Hint}) and read
\begin{eqnarray}
  \label{eq:init_cond_RG}
  A^\rmc_\gamma(\Lambda_0) &=& B^\rmc_\gamma(\Lambda_0) 
  = C^\rmc_\gamma(\Lambda_0) = U(\delta_{\gamma,1}+\delta_{\gamma,-1})
  \nonumber\\
  A^\rms_\gamma(\Lambda_0) &=& B^\rms_\gamma(\Lambda_0) 
  = C^\rms_\gamma(\Lambda_0) = -U(\delta_{\gamma,1}+\delta_{\gamma,-1})
  \nonumber\\
  D^\rmc(\Lambda_0) &=& 2U\\
  D^\rms(\Lambda_0) &=& -2U\nonumber
\end{eqnarray}
The interactions are renormalized both by particle-hole and particle-particle 
contributions.
Written with the dimensionless scale $l=\ln(\Lambda_0/\Lambda)$, the 
RG equations are
\begin{widetext}
  \begin{eqnarray}
    \label{eq:RG_equations_A}
    \pal A^\rmcs_\gamma &=& 
    \Big[ A_\gamma * A_\gamma 
    + C_\gamma * C_{-\gamma}\Big]_\rmph^\rmcs
    + \Big[ A_\gamma *A_\gamma 
    + B_\gamma * B_{-\gamma}\Big]_\rmpp^\rmcs,\\
    \label{eq:RG_equations_B}
    \pal B^\rmcs_\gamma &=& 
    \Big[ 2 D * B_\gamma 
    + \sum_{\beta\neq 0,\gamma} B_\beta * B_{\gamma-\beta}\Big]_\rmph^\rmcs
    + \Big[ B_\gamma * (A_\gamma + A_{-\gamma})\Big]_\rmpp^\rmcs,\\
    \label{eq:RG_equations_C}
    \pal C^\rmcs_\gamma &=& 
    \Big[ C_\gamma * (A_\gamma + A_{-\gamma})\Big]_\rmph^\rmcs
    + \Big[ 2 D * C_\gamma 
    + \sum_{\beta\neq 0,\gamma} C_\beta * C_{\gamma-\beta}\Big]_\rmpp^\rmcs,\\
    \label{eq:RG_equations_D}
    \pal D^\rmcs &=& 
    \Big[ D*D 
    + \sum_{\beta\neq 0} B_\beta * B_{-\beta}\Big]_\rmph^\rmcs
    + \Big[ D*D 
    + \sum_{\beta\neq 0} C_\beta * C_{-\beta}\Big]_\rmpp^\rmcs .
  \end{eqnarray}
\end{widetext}
In these flow equations, we have denoted
\begin{eqnarray}
  \Big[G_1 * G_2 \Big]_\rmph^\rmc &=& \frac{1}{2\pi} \Big( G_1^\rmc G_2^\rmc 
  + 3 G_1^\rms G_2^\rms \Big),\\
  \Big[G_1 * G_2 \Big]_\rmph^\rms &=& \frac{1}{2\pi} \Big( G_1^\rms G_2^\rmc 
  + G_1^\rmc G_2^\rms + 2 G_1^\rms G_2^\rms \Big),\quad
\end{eqnarray}
for the particle-hole channel, and
\begin{eqnarray}
  \Big[G_1 * G_2 \Big]_\rmpp^\rmc &=& - \frac{1}{2\pi} \Big( G_1^\rmc G_2^\rmc 
  + 3 G_1^\rms G_2^\rms \Big),\\
  \Big[G_1 * G_2 \Big]_\rmpp^\rms &=& - \frac{1}{2\pi} \Big( G_1^\rms G_2^\rmc
  + G_1^\rmc G_2^\rms - 2 G_1^\rms G_2^\rms \Big),\quad
\end{eqnarray}
for the particle-particle channel.
The RG flow equations can be simplified further, but we shall leave them 
in the form (\ref{eq:RG_equations_A}-\ref{eq:RG_equations_D}) because it 
clearly shows the origin of all one-loop contributions. As an illustration, 
we show in Fig.~\ref{fig:renC} the bubbles renormalizing the pair hopping term 
$C_\gamma$.

%


\subsection{Numerical solution of the Renormalization Group equations}

\subsubsection{Strong coupling Hamiltonian}

The flow equations (\ref{eq:RG_equations_A}-\ref{eq:RG_equations_D}) cannot
be solved analytically, but one can integrate them numerically with standard
numerical routines such as Runge-Kutta of order 4. One nice feature of the
local $\mathbb Z_2$-symmetry is that it constrains interactions in such a way that 
they can depend on at most one cage index. Thus the number of equations that have to 
be solved only grows linearly with the total number of cages $N$, allowing 
to consider large system sizes. 

With the initial conditions (\ref{eq:init_cond_RG}), we found that the 
system goes to strong coupling. The flow diverges at a critical RG scale 
$l_c$, indicating a phase transition at a critical temperature of order of
magnitude $T_\rmc=\Lambda_0\exp(-l_c)$. Apart from finite-size corrections, 
the numerical solution of the flow reaches the following fixed direction
\begin{eqnarray}
  \label{eq:fixed_dir}
  A^\rmc_\gamma &=& A^\rms_\gamma=0,\\
  B^\rmc_\gamma &=& D^\rmc = -3 B^\rms_\gamma = -3 D^\rms>0,\\
  C^\rmc_\gamma &=& C^\rms_\gamma=0,
\end{eqnarray}
for any value of $\gamma$. The corresponding effective
interaction Hamiltonian can be written (going back to real space and setting
the strength of the interaction to a value $g$)
\begin{eqnarray}
  \label{eq:effective_ham}
  H_\mathrm{int}^\mathrm{eff} &=& g \int_{-\infty}^\infty \rmd x 
  \sum_{n,\gamma} :\Big[ 
  3 \rho_{\rmr,n+\gamma,n}(x) \rho_{\rml,n,n+\gamma}(x)\nonumber\\
  && - \bS_{\rmr,n+\gamma,n}(x) \bS_{\rml,n,n+\gamma}(x).
\Big]:
\end{eqnarray}
This Hamiltonian has an SU($N$) symmetry in the cage indices, and it
corresponds to a particle-hole pairing in the triplet channel. Consequently 
the system undergoes a phase transition to a spin-density wave state, as was
predicted by our mean-field analysis (see appendix~\ref{AppSDW}).
This is more transparent on the equivalent form of~(\ref{eq:effective_ham}):
\begin{eqnarray}
  \label{eq:effective_ham_newlook}
\lefteqn{H_\mathrm{int}^\mathrm{eff}=-2g \int_{-\infty}^\infty \rmd x 
  \sum_{m,n}\sum_{\mu\mu'}\sum_{\nu\nu'}}\\
& & :\left(\cc_{\mathrm{R},m,\mu}(x)\boldsymbol{\sigma}_{\mu,\mu'}\ac_{\mathrm{L},m,\mu'}(x)\right)
\left(\cc_{\mathrm{L},n,\nu}(x)\boldsymbol{\sigma}_{\nu,\nu'}\ac_{\mathrm{R},n,\nu'}(x)\right):\nonumber 
\end{eqnarray}
since it is bilinear in the $2k_{\mathbf{F}}$ Fourier component of the local spin density.
A more detailed analysis of this phase, including a discussion of its
collective modes can be found in Ref.~\onlinecite{Dusuel02}.

The RG calculation however gives a much smaller value of the critical
temperature than the mean-field calculation. We have computed the critical
temperature given by the full flow equations 
(\ref{eq:RG_equations_A}-\ref{eq:RG_equations_D}), as well as for the RPA
approximation of these, which consists in neglecting all particle-particle
contributions. We found the following numerical result
\begin{equation}
  \label{eq:critical_temp}
  l_\rmc(\mathrm{RG}) \simeq \frac{1.33}{U} \quad\mbox{ and }\quad
  l_\rmc(\mathrm{RPA}) \simeq \frac{0.64}{U}.
\end{equation}
For an initial interaction $U=0.1$, this gives a ratio 
$T_\rmc(\mathrm{RG})/T_\rmc(\mathrm{RPA})\simeq 10^{-3}$, which is very small. 
The RPA thus overestimates the critical temperature by three orders of 
magnitude for $U=0.1$. For the typical value $\alpha=0.4$ $(U=2\pi\alpha)$ estimated in
GaAs networks, this reduction effect is not as dramatic, since here we
obtain: $T_\rmc(\mathrm{RG})/T_\rmc(\mathrm{RPA})\simeq 0.76$. This gives further
support to replacing finite Aharonov-Bohm cages by infinite ones in  
our simplified model, since the energy scale which emerges from the RG treatment is still
in the 1D regime, for a typical choice of parameters such as $L/\lambda_{\mathrm{F}}=40$.

\subsubsection{Intermediate energy regime}

Another feature of the RPA is that the pair hopping terms $C^\rmcs$ remain
nearest neighbor all along the flow, as can be inferred from 
Eq.~(\ref{eq:RG_equations_C}), when neglecting the particle-particle
contribution. This is obviously not the case any more for the full RG
equations. As the pair hopping terms are the only processes that can lead to a
conducting behavior, it is natural to study what happens to them during the
flow, before they get completely suppressed by the spin-density wave.

\begin{figure}[t]
  \centering
  \includegraphics[width=8cm]{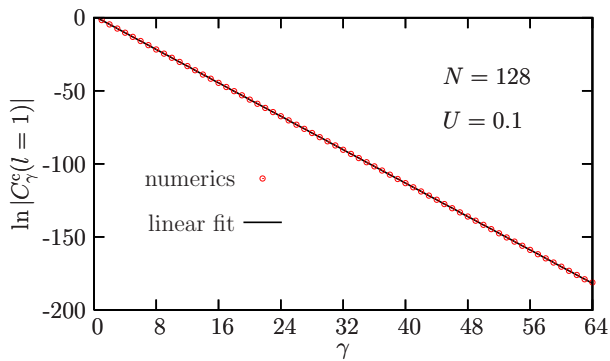}
  \caption{The numerical results of $\ln|C_\gamma^\rmc(l=1)|$ as a
  function of $\gamma$, for $N=128$ cages and initial interaction $U=0.1$, 
  are perfectly fitted by a straight line.}
\label{fig:fitexp}
\end{figure}

\begin{figure}[t]
  \centering
  \includegraphics[width=8cm]{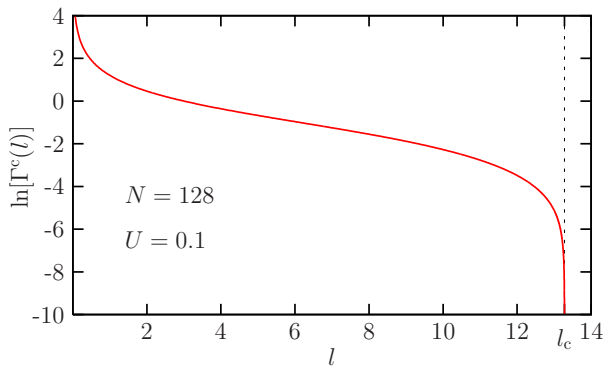}
  \caption{Evolution of the strength of the interaction $C^\rmc_\gamma(l)$ as a
  function of the RG scale $l$, for $N=128$ cages and initial interaction
  $U=0.1$. The flow stops at the critical scale $l_\rmc$ denoted by a dashed
  line.}
\label{fig:valueCc}
\end{figure}

\begin{figure}[t]
  \centering
  \includegraphics[width=8cm]{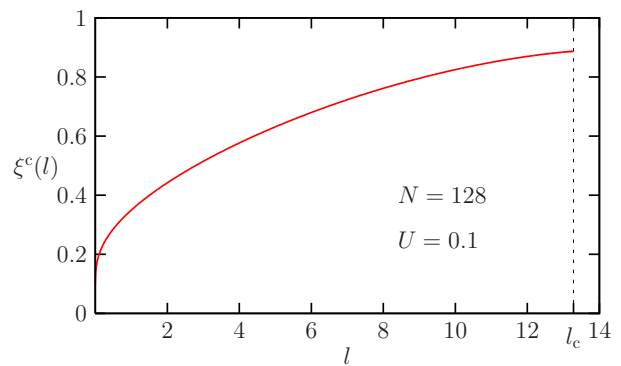}
  \caption{Evolution of the range of the interaction $C^\rmc_\gamma(l)$ as a
  function of the RG scale $l$, for $N=128$ cages and interaction strength
  $U=0.1$. The flow stops at the critical scale $l_\rmc$ denoted by a dashed
  line.}
\label{fig:rangeCc}
\end{figure}

We found numerically that, except at the very beginning of the flow, the pair
hopping terms are very well fitted by the following law
\begin{equation}
  \label{eq:fitC}
  C_\gamma^\rmcs(l) = (-1)^\gamma \Gamma^\rmcs(l) \exp[-\gamma/\xi^\rmcs(l)].
\end{equation}
This shows the incoherent nature of these pair hopping processes.
As an illustration, Fig.~\ref{fig:fitexp} shows $\ln|C_\gamma^\rmc(l=1)|$ as a
function of $\gamma$, for $N=128$ cages and interaction strength $U=0.1$.
We have performed such fits for many values of $l$ so as to extract the scale
dependence of the quantities $\Gamma^\rmcs(l)$ and $\xi^\rmcs(l)$ defined in
Eq.~(\ref{fig:fitexp}). The results for the charge and spin couplings are 
nearly identical, so that we only show the behavior of the charge quantities, 
see Figs.~\ref{fig:valueCc} and \ref{fig:rangeCc}.
From Fig.~\ref{fig:valueCc}, it is clear that the pair hopping is suppressed 
during the flow, but there is an intermediate energy regime 
($2\lesssim l \lesssim 12$) where the decrease rate of the pair hopping is not
too large. Furthermore, we see in Fig.~\ref{fig:rangeCc} that the typical
range of the pair hopping increases during the flow, but it does not 
exceed one lattice spacing. The pair hopping thus remains local, contrarily to the
cage exchange terms $B_\gamma$ which become of infinite range during the flow,
as was explained in the previous subsection.

\subsection{Validity of the simplified model}

An important question arises: when can the Friedel density oscillations be ignored?
In a real system, two physical processes are in competition: the wire disconnection at network junctions due to the $S$-matrix renormalization, and the SDW instability discussed earlier. 
In order to estimate the typical energy scale of wire disconnection, we use Eq.~(\ref{scalingT}) 
and take the point where both terms in the denominator are of the same order of magnitude. 
\begin{equation}
k_\mathrm{B}T_\mathrm{c}(\mathrm{S})\simeq E_\mathrm{F}\left(\frac{R_0}{T_0}\right)^{1/2\alpha}
\end{equation}
The simplified model discussed in this section gives a faithful description of the initial
$\mathbb Z_2$ network only if the typical energy scale for the SDW instability is larger
than $T_\mathrm{c}(\mathrm{S})$. From~(\ref{eq:critical_temp}), we have:
\begin{equation}
k_\mathrm{B}T_\mathrm{c}(\mathrm{RG})\simeq E_\mathrm{F}\exp\left(-\frac{1.33}{U}\right)
\end{equation}
This gives a rather precise criterion, namely:
\begin{equation}
\label{valid_criterion}
\frac{R_0}{T_0}<\exp(-\frac{1.33}{\pi})\simeq 0.655
\end{equation}
Note that because of the one loop RG approximation used for both mechanisms (i.e
scattering on Friedel oscillations, and SDW instability), this criterion is independent
of the interaction strength, provided it is small compared to the electronic Fermi energy.
We may say that it is of a purely geometric nature, since it mostly depends on the
amount of electronic backscattering induced by the network nodes. For an illustration, let us
consider the two types of node geometries depicted on Fig.~\ref{4node}.
For single mode wires, we have found that the lowest values of the ratio $R_0/T_0$ 
for perfectly symmetric $\mathrm{X}$ or $\mathrm{Y}$ junctions are equal respectively to $1$ and $1/4$.
So changing the junction type from $\mathrm{X}$ (Fig.~\ref{4node} a)) to $\mathrm{Y}$ \mbox{(Fig.~\ref{4node} b))} 
could affect deeply the nature of the low-energy state.

In the case of a single 1D quantum wire, the RG approach
gives a vanishing transition temperature and the spin density wave is
replaced by a Luttinger liquid fixed point at low energy.
For $\mathbb Z_2$ networks in the regime where $T_{\mathrm{c}}(\mathrm{RG})>T_{\mathrm{c}}(\mathrm{S})$,
inter-cage interaction processes destroy
this Luttinger liquid, and because of the flow to strong coupling,
most elementary excitations acquire a gap of the order $T_\rmc(\mathrm{RG})$.
In the other regime, ($T_{\mathrm{c}}(\mathrm{RG})<T_{\mathrm{c}}(\mathrm{S})$),
our simplified model discarding Friedel oscillations cannot capture the low energy physics
below  $T_{\mathrm{c}}(\mathrm{S})$. Below this scale, the network ``breaks'' into independent
disconnected links, which are expected to behave as independent finite segments of
Luttinger liquids.

How could we distinguish between these two situations in real experiments?
It may not be very easy at a qualitative level, just looking at 
transport measurements. Resistivity is expected to rise sharply below the
largest temperature scale between $T_{\mathrm{c}}(\mathrm{RG})$ and $T_{\mathrm{c}}(\mathrm{S})$, but 
for very different reasons in each case. If $T_{\mathrm{c}}(\mathrm{RG})>T_{\mathrm{c}}(\mathrm{S})$, the
single electron spectrum is gapped, and charge transport is provided only via a single 
charge collective mode, which is likely to be quite sensitive to pinning by random impurities.
If $T_{\mathrm{c}}(\mathrm{RG})<T_{\mathrm{c}}(\mathrm{S})$, the disconnection of the network into independent
segments clearly induces a very high resistivity. The difference between the two situations
could be evidenced by {\em non-linear} dc-transport. Indeed, in the former situation,
a pinned spin-density wave can be depinned by a large enough electric field, and should
lead to a large non-linear dc conductivity above a finite threshold value for the driving field.
In the later case, possible non-linear effects on the disconnection mechanism have not
been investigated to our knowledge. But it seems unlikely that they would yield a sharp
threshold as in the case of collective depinning.

The above discussion shows that it would be crucial to estimate the ratio
$R_0/T_0$ for a given sample. In present experiments, although the number
of channels $N_\mathrm{ch}$ is low (about five), it is not exactly one. 
So we have a $Z\times N_{ch}$ conduction matrix at each node connected to $Z$
nearest neighbors. This clearly calls for an extension of the theory
to the case of multichannel wires, which goes beyond the scope of the present work.
But even for a strictly single channel network, a direct experimental determination of 
$R_0/T_0$ may not be straightforward. In a dc transport measurement of an {\em ideal}
system, the Landauer approach shows that the resistivity is dominated by the contacts
between the network and the outer current leads. As shown in~\cite{Kazymyrenko},
$R_0/T_0$ controls the dispersion of the mini energy bands associated to the
lattice superstructure. If $R_0/T_0$ is small, these bands are weakly dispersive,
and the effect of static random impurities, always present in a real system,
is stronger than for a large value of $R_0/T_0$. But in general, we do not have
an independent measurement of the disorder strength. So at this stage, the main
prediction is that some $\mathbb Z_2$ networks should develop a SDW instability
and some others break into independent disconnected segments at low energy.
Although the mathematical criterion~(\ref{valid_criterion}) is quite precise and
simple, it may be difficult {\em in practise} to decide if a given sample
will fall in one or the other class without actually doing the experiment.


\section{Conclusion}

In this paper, we have studied a class of networks ($\mathbb Z_2$ networks)
for which all single electron energy eigenstates are localized in Aharonov-Bohm 
cages when an external magnetic field corresponding to half a flux quantum per 
loop is applied. The particularity of these networks (in comparison to other geometries
showing Aharonov-Bohm cages such as the dice lattice), is the presence of a large
group of local $\mathbb Z_2$ symmetries, which correspond to reversing the electronic
current in all the loops adjacent to a given node. Interestingly, these symmetries
are preserved in the presence of a large class of interaction potentials, including 
the Hubbard local interaction. The conserved quantum numbers associated to this
symmetry are simply the parity of the total electron number in each Aharonov-Bohm cage.
Because of this strong constraint, only pairs of electron can tunnel from one cage to
another. 

A first consequence, assuming that the local $\mathbb Z_2$ symmetry cannot be spontaneously
broken, is that such a system is {\em never} a Landau Fermi liquid, since the
single electron propagator is short ranged in space. This establishes a qualitative
difference between $\mathbb Z_2$ networks and geometries such as the dice lattice.
In the latter case, Aharonov-Bohm cages are no longer protected by a symmetry in the
interacting system, and therefore a Fermi liquid induced by interactions seems
perfectly plausible. This extends the distinction established for bosonic systems
for which a single boson condensate is always destroyed on a $\mathbb Z_2$ network,
whereas it survives on the dice lattice in spite of a large discrete ground-state degeneracy. 

Using a simple mean-field analysis, completed by an unbiased renormalization group (RG)
approach, we first conclude that no superconducting instability is expected, in spite
of the particle pair hopping term generated by electron-electron interactions
preserving the local $\mathbb Z_2$ symmetry. The RG analysis shows that long-ranged
pair hoppings are generated at intermediate energies, but the typical hopping
range remains of the order of the lattice spacing. This time, these results
are in sharp contrast with the bosonic counterpart, where the absence of a single
boson condensate leaves room for a perfectly stable boson {\em pair} condensate~\cite{Doucot02},
with very interesting properties such as topological ground-state degeneracies~\cite{Ioffe02}.
The main result given by the RG is the presence of a strong coupling low-energy fixed point,
dominated by the long-distance fluctuations of a local spin density wave order parameter.
The appearance of this fixed point can be attributed to the presence of Aharonov-Bohm cages,
since in the model considered here, independent links would behave as Luttinger
liquids.

We should note that the RG calculation has been done on a simplified model, where all
the lattice points inside a given Aharonov-Bohm cage are equivalent. So we have
not incorporated the physics of Friedel density oscillation induced by the nodes,
which are known to renormalize the single-electron scattering matrices at these nodes,
as long as the system is in the one dimensional regime above the energy scale $hv_{\mathrm{F}}/L$,
which can be as large as a few Kelvin degrees in a GaAs/GaAlAs structure.
The validity range of this simplified model has been formulated in terms of
a very simple criterion on the value of the reflexion coefficient at the network
nodes. It has to be low enough, otherwise, the disconnection induced
by electronic scattering on Friedel density oscillations dominates over
the tendency to build a SDW instability, and then, the low energy state
appears to be a collection of independent Luttinger liquid segments.

This study raises some open questions regarding the transport properties
of such incoherent $\mathbb Z_2$ metal. It would be very interesting to
compute the temperature dependence of the conductivity. At temperatures
above the spin density wave instability, it is likely to exhibit a power-law
behavior, with a presently unknown exponent. In the strong coupling regime,
a gap for single electronic excitations is expected, so most likely,
the temperature dependence of the conductivity will show an activated behavior.
Another fascinating question is the transport through hybrid structures
such as a series of a normal metal, a $\mathbb Z_2$ network, and a normal metal.  
Indeed, for driving voltages below the $\mathbb Z_2$ gap (of the order of $hv_{\mathrm{F}}/L$),
incoming (resp. outgoing) electrons will have a strong tendency to be injected
(resp. emitted) in pairs, although the $\mathbb Z_2$ network is far from
a superconducting state. Maybe the most direct way to demonstrate
this strange phenomenon in such a poorly conducting system would be through
shot noise measurements.


\acknowledgments

We wish to thank J. Dufouleur, G. Faini, and D. Mailly for giving us 
the motivation for this work, and for several useful discussions.
We also thank J. Vidal for his comments on the manuscript.  
Financial support of the DFG in SP1073 is gratefully acknowledged.

\appendix

\section{Spin density wave instability on a $\mathbb Z_2$ network}
\label{AppSDW}

For the sake of simplicity, let us consider the Hubbard local interaction.
For an infinite 1D system, we decouple the interaction as:
\[
U\left(\langle\cpsi_{\uparrow}(x)\apsi_{\uparrow}(x)\rangle\cpsi_{\downarrow}(x)\apsi_{\downarrow}(x)
+\langle\cpsi_{\downarrow}(x)\apsi_{\downarrow}(x)\rangle\cpsi_{\uparrow}(x)\apsi_{\uparrow}(x)\right)
\]
Let us assume that \mbox{$\langle\cpsi_{\tau}(x)\apsi_{\tau}(x)\rangle=\tau\Delta\cos(2k_{\mathrm{F}}x)$}.
In linear response theory, self-consistency determines the mean-field critical
temperature $T^{*}$ from the requirement \mbox{$U\chi^{(0)}(2k_{\mathrm{F}},T^{*})=1$},
where $\chi^{(0)}(q,T)$ is the static charge susceptibility of the non-interacting
electron gas at wave-vector $q$ and temperature $T$. Because
\mbox{$\chi^{(0)}(2k_{\mathrm{F}},T)=(hv_{\mathrm{F}})^{-1}\ln(aE_{\mathrm{F}}/k_{\mathrm{B}}T)$},
where $a$ is a numerical factor of order unity, the mean-field temperature $T^{*}$
obeys a BCS relation:
\begin{equation}
k_{\mathrm{B}}T^{*}=aE_{\mathrm{F}}\exp(-\frac{hv_{\mathrm{F}}}{U})
\label{meanfieldTc}
\end{equation} 

For a $\mathbb Z_2$ network, our starting point is Eq.~(\ref{hubbard}). Again, we
are looking for a $\mathbb Z_2$ invariant order parameter of the form:
\mbox{$\langle\cpsi_{i\tau}(x)\apsi_{i\tau}(x)\rangle=\tau\Delta_{i}(x)$}.
With such an order parameter, the quantum expectation value of Eq.~(\ref{hubbard})
taken for a Hartree-Fock state is equal to \mbox{$-U(\Delta_{i}(x)+\Delta_{j}(x))^{2}$}.
We see that choosing a uniform spin density wave order parameter \mbox{$\Delta_{i}(x)=\Delta$}
definitely lowers the system interaction energy, and we expect a mean-field transition
temperature of the same order of magnitude as for the single infinite wire with a repulsive
local interaction. Note that this transition is uniquely driven by the first two
terms in Eq.~(\ref{hubbard}), that is the intra and inter cage interaction. The last two
terms, i.e. the cage exchange and the pair hopping processes do not play a role in 
stabilizing the spin density wave state. As seen in section~\ref{RG} in a more elaborate
treatment, these last two terms have the effect to lower the critical temperature
by comparison to the mean-field estimate.


\end{document}